\documentclass[journal]{vgtc}         

\ifpdf
  \pdfoutput=1\relax                   
  \usepackage{graphicx}                
  \DeclareGraphicsExtensions{.pdf,.png,.jpg,.jpeg} 
\else
  \usepackage{graphicx}                
  \DeclareGraphicsExtensions{.eps}     
\fi%

\graphicspath{{figures/}{pictures/}{images/}{./}} 

\usepackage{microtype}                 
\PassOptionsToPackage{warn}{textcomp}  
\usepackage{textcomp}                  
\usepackage{mathptmx}                  
\usepackage{times}                     
\usepackage{cite}                      
\usepackage{tabu}                      
\usepackage{booktabs}                  
\usepackage{xcolor}
\usepackage{amssymb}
\usepackage{soul}
\usepackage{enumitem}

\onlineid{1011}

\vgtccategory{Research}
\vgtcpapertype{Algorithm/Technique}

\title{Persistence Atlas for Critical Point Variability in 
Ensembles}

\author{
Guillaume Favelier,
Noura Faraj,
Brian Summa, and
Julien Tierny}
\authorfooter{
\item G. Favelier and J. Tierny are with Sorbonne Universit\'e and CNRS 
(LIP6).\\ 
E-mail: \{guillaume.favelier, julien.tierny\}@sorbonne-universite.fr.
\item N. Faraj and B. Summa are with Tulane University. \\
E-mail: 
\{nfaraj, bsumma\}@tulane.edu.
}

\shortauthortitle{Favelier \MakeLowercase{\textit{et al.}}: 
Persistence Atlas for Critical Point Variability in Ensembles}

\abstract{
This paper presents a new approach for the visualization and analysis of the 
spatial variability  of features of interest represented by 
critical points in ensemble data. 
Our framework, called \emph{Persistence Atlas}, enables the 
visualization of the dominant spatial patterns of critical points,
along with statistics regarding their occurrence in the 
ensemble. 
The persistence atlas represents in the geometrical domain each dominant 
pattern in the form of a confidence map for the appearance of critical points.
As a by-product, our method also provides 2-dimensional layouts of the 
entire ensemble, highlighting the main trends at a global level. 
Our approach is based on the new notion of \emph{Persistence Map}, 
a measure of the geometrical density in critical points which leverages the 
robustness to noise of topological persistence to better emphasize salient 
features. We show how to leverage spectral embedding to represent the ensemble 
members as points in a low-dimensional Euclidean space, where distances between 
points measure the dissimilarities between critical point layouts and where 
statistical tasks, such as clustering, can be easily carried out. 
Further, we show how the notion of 
mandatory critical point can be leveraged to evaluate for each cluster 
confidence regions for the appearance of critical points. Most of the steps of 
this framework can be trivially parallelized and we show how to efficiently 
implement them. 
Extensive experiments 
demonstrate the relevance 
of our approach. The accuracy of the confidence regions provided by the 
persistence atlas is quantitatively evaluated and compared to a baseline 
strategy using an off-the-shelf clustering approach. We illustrate the 
importance of the persistence atlas in a variety of real-life datasets, where 
clear trends in feature layouts are identified and analyzed.
We provide a lightweight VTK-based C++ implementation of our approach that can 
be used for 
reproduction purposes.

} 


\CCScatlist{ 
 \CCScat{K.6.1}{Management of Computing and Information Systems}%
{Project and People Management}{Life Cycle};
 \CCScat{K.7.m}{The Computing Profession}{Miscellaneous}{Ethics}
}

\newcommand{\vcaption}[1]{
\vspace{-1ex}\caption{#1}\vspace{-1.5ex}}

\teaser{
  \centering
  \includegraphics[height=4.85cm]{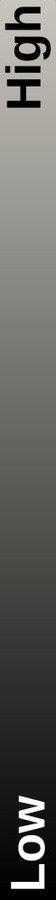}
  \hfill
  \includegraphics[height=4.85cm]{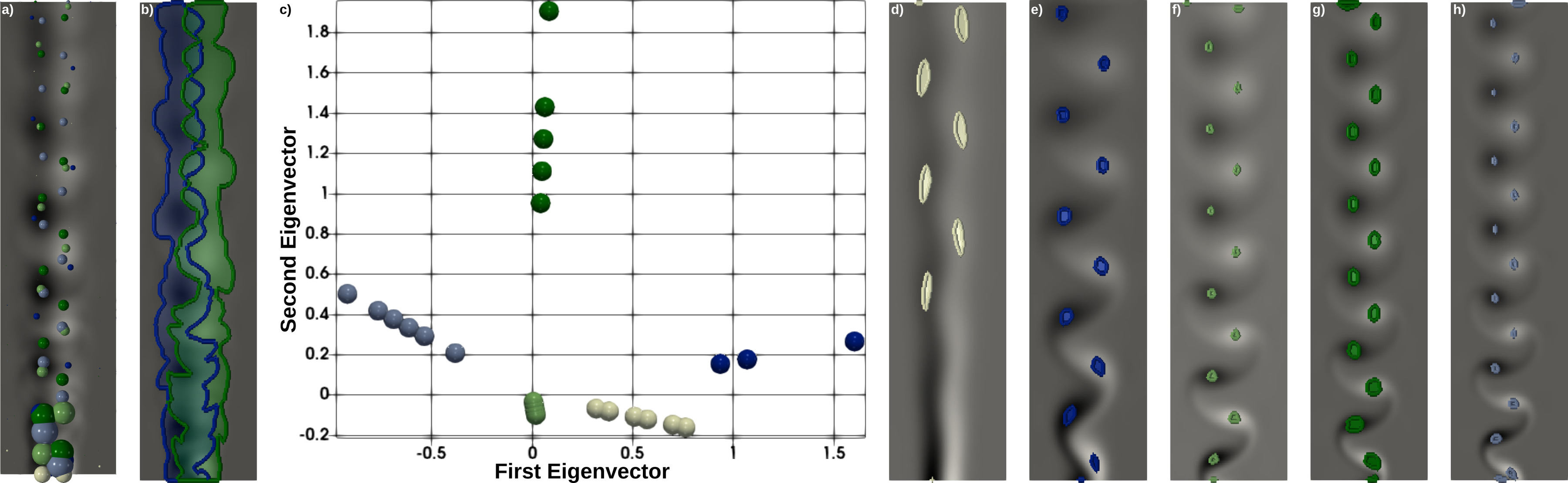}
  \vcaption{
  Persistence atlas for an ensemble of 45 von K\'arm\'an vortex streets 
  (scalar data: orthogonal component of the curl).
  (a) Critical points 
  (minima and maxima,
  scaled by persistence) of a few representative ensemble 
members (one 
color per member) 
exhibit 
clearly distinct layout patterns in terms of position 
and number of vortices, revealing high \emph{spatial} and \emph{trend}
variabilities within the ensemble. (b) Mandatory critical points
(minimal regions where at least one critical point is guaranteed 
to occur for every member of the ensemble) are thus particularly 
conservative given these variabilities and identify only one region per side of 
the vortex street (blue: minimum, green: maximum).
(c) The persistence atlas addresses this issue by analyzing the structure of 
the ensemble in terms of critical point layouts and provides low dimensional 
embeddings of the members where statistical tasks, such as clustering, can be 
easily carried out. 
In particular, our approach  automatically identified  
five clusters, 
(d) to (h), 
corresponding to five 
distinct trends in critical point layouts (five  
viscosity regimes). Per cluster mandatory critical points 
provide more accurate and useful critical point predictions (colored regions, 
(d) to (h)), revealing 
an increasing number of vortices and a
decreasing spatial variability for increasing Reynolds 
numbers (left to right). The background color map shows the mean scalar field 
for the 
entire ensemble, (a) and (b), and individual clusters, (d) to (h).
\vspace{2ex}
}
	\label{fig_teaser}
}

\vgtcinsertpkg

\newcommand{\julien}[2]{\textcolor{black}{#2}}
\newcommand{\brian}[2]{\textcolor{black}{#2}}

\newcommand{\BrianSays}[1]{\textcolor{blue}{***BRIAN: #1***}}
\renewcommand{\BrianSays}[1]{}

\newcommand{\domain}{\mathcal{M}}

\newcommand{\stt}[1]{St(#1)}
\newcommand{\simplex}{\sigma}
\newcommand{\lk}[1]{Lk(#1)}
\newcommand{\lkminus}[1]{Lk^-(#1)}
\newcommand{\lkplus}[1]{Lk^+(#1)}
\newcommand{\sub}[1]{f^{-1}_{-\infty}(#1)}

\newcommand{\Index}{\mathcal{I}}

\newcommand{\persistenceDiagram}[1]{\mathcal{D}(#1)}
\newcommand{\equref}[1]{Eq.~\ref{#1}}

\begin{document}


\firstsection{Introduction}

\maketitle

In engineering and 
science, measurements and simulations 
are necessary to 
understand complex physical systems (in chemistry, 
astrophysics, climate modeling, weather forecasts, etc.) and, 
more importantly, to 
try to predict their behavior. 
Modern simulations are subject to a variety of input parameters, 
related to the initial conditions of the system or to the configuration of its 
environment. Given the increase in computational power typically observed with 
current supercomputers and high-end workstations, it is now possible for 
engineers and scientists to densely sample the space of input
parameters to 
compute \emph{ensembles} formed from many numerical 
simulations, in order to better 
understand the
variability of the system with regard to its input parameters. 
In the case of scalar variables, this means that the data which is 
considered for 
visualization and analysis is no longer a single field, but an ensemble
of scalar fields, for which global trends or outliers need to be identified, 
visualized and analyzed. \BrianSays{Previous two sentences need to be 
reworked.  Too long} For measured data, the inherent uncertainty 
of the acquisition process can also be represented in terms of ensembles, by 
considering realizations of the random variable characterizing the acquisition 
noise for each point of the domain. \BrianSays{I don't like how the 
first sentence goes: measurement and simulations, but the paragraph is 
simulations then measurements}


Ensemble datasets are however notoriously difficult to analyze, visualize and 
interpret. First, features of interest in ensemble data 
exhibit \emph{(i)} \emph{trend variability}: distinct 
trends often
emerge among groups of ensemble members which 
share a similar configuration 
in terms of the number and location of features of interest. Second, even 
when ensemble members share common trends, features are affected by \emph{(ii)}
\emph{spatial variability}: their exact location may
vary among the ensemble members. 
Both types of 
variabilities must be analyzed, 
quantified, and visualized to \brian{help}{aid} \brian{users}{users'} 
\brian{apprehend}{understanding of} 
the structure of their 
ensemble data and better predict 
the behavior of the system \brian{under consideration}{} in terms of the possible configurations of 
features of interest. 
Taken individually, the ensemble members may not be 
representative of the major trends in the  whole ensemble. Hence, their 
direct 
visualization does not account for trend variability.
Moreover, they are often 
too numerous to allow interactive inspection. 
In contrast, naive aggregation 
measures, such as point-wise means, drastically smooth details 
out, 
preventing the identification of features of interest 
 which only occur in subsets of the ensemble members or 
with high spatial variability.

Thus, 
it is necessary to introduce advanced techniques for the analysis of the 
features of interest in ensemble data, to 
\emph{(i)} identify the trends in feature configurations,
\emph{(ii)} estimate their respective appearance statistics,
and to \emph{(iii)} characterize  their respective spatial
variability.
While this overall strategy has been successfully instantiated for 
simple \brian{geometrical }{ }objects, such as level sets \cite{WhitakerMK13, 
FerstlKRW16} 
or streamlines \cite{FerstlBW16}, it is necessary to extend it to
more advanced constructions, such as topological features.
Topological data analysis (TDA) \cite{edelsbrunner09, heine16} has
demonstrated its ability over the last two decades to capture in a generic, 
robust and efficient manner the features of interest in scalar data in a 
variety of applications: turbulent combustion \cite{laney_vis06, 
bremer_tvcg11, gyulassy_ev14}, material sciences \cite{gyulassy_vis07, 
gyulassy_vis15, favelier16}, computational fluid dynamics \cite{kasten_tvcg11}, 
chemistry \cite{chemistry_vis14, harshChemistry} or astrophysics 
\cite{sousbie11, shivashankar2016felix} to name a few. In these
applications, domain-specific features of interest are easily expressed in 
terms of the critical points \cite{banchoff70} of the data (points where the 
gradient vanishes), which are 
robustly extracted by topological methods. 
\BrianSays{suggest removing the last part of the previous sentence. Extraneous} 
For instance, critical points 
capture atomic structures in molecular chemistry \cite{chemistry_vis14, 
harshChemistry}, flame centers in combustion \cite{bremer_tvcg11, 
gyulassy_ev14}, vortices in fluid dynamics \cite{kasten_tvcg11}, etc. 
However, despite their importance 
in applications, the trend variability of critical points in ensemble data has 
not been investigated so far.

This paper fills this gap with the concept of \emph{Persistence Atlas}, a global
approach for critical point variability analysis in ensembles of scalar fields. 
This approach is based on the new notion of \emph{Persistence Map}, 
which characterizes for each member of the ensemble the 
spatial distribution 
of its critical points. To be robust to the presence of noise in the data, this 
representation 
takes into account for each 
critical point its topological persistence \cite{edelsbrunner02}, an established 
measure of importance in topological data analysis. 
Based on 
this persistence map representation,
our approach  
embeds each member of the ensemble as a point in a low-dimensional Euclidean 
space, where the distance between two members measures the similarity between 
their critical point distributions. 
This embedding is 
exploited to derive a 
global representation
of the distributions of critical points 
within the ensemble, allowing for the automatic identification of clusters, 
revealing the major trends in critical point layouts in the ensemble. 
Additionally, for each cluster, we show how the notion of mandatory critical 
point \cite{tierny_ev14} can be used to derive relevant confidence regions for 
the appearance of critical points in the spatial domain. 
The \brian{output}{} persistence atlas is then a composition of 
a series of confidence maps 
for the appearance of salient critical points, each map revealing a specific 
trend of the ensemble.

Extensive experiments on synthetic and real-life data demonstrate the relevance 
of our persistence map representation for the comparison of critical point 
distributions among ensemble members. The clustering performance of our 
framework
and the accuracy of  its confidence 
regions 
are
quantitatively evaluated and 
compared to a baseline strategy using an off-the-shelf clustering approach. We 
illustrate the importance of the persistence atlas for a number of 
real-life 
datasets, where clear trends in feature layouts are identified and analyzed.



\subsection{Related work}
\label{sec_relatedWork}
The literature related to our approach can be classified into two categories: 
\emph{(i)} uncertainty visualization and \emph{(ii)} ensemble visualization. In 
the first case, the data variability is explicitly encoded by an estimator of 
the probability density function (PDF) of a point-wise random variable (usually 
with strong assumptions on its structure)\brian{, while i}{. I}n the second, the variability 
arises from a series of global empirical observations.

\noindent
\textbf{(i) Uncertainty visualization:} The analysis and visualization of 
uncertainty in data \cite{GUM} is commonly recognized as an important 
yet long standing challenge in the visualization community \cite{Pang97, 
johnson03,
MacEachren05}, as documented in several surveys \cite{potter12, 
bonneau:hal-01060465}. Given an estimator of the PDF of a random variable 
modeling the point-wise variability in data values (for instance, a Gaussian 
distribution), several techniques have been proposed to represent the 
distribution of the uncertainty in the data, by either considering the entropy 
of the random variables \cite{potter13}, correlation in uncertainties 
\cite{DBLP:journals/cgf/PfaffelmoserW12} or variability in the data 
gradient \cite{DBLP:journals/tvcg/PfaffelmoserMW13}.\BrianSays{prev sentence too long} To understand the 
positional uncertainty related to \brian{geometrical}{geometric} constructions generated out of 
the raw uncertain data, specialized methods have been designed\brian{, f}{. F}or \brian{instance}{example,} \brian{to 
estimate}{approaches have estimated} the positional uncertainty of level sets under 
various interpolation schemes and PDF models \cite{pfaffelmoser11, poetkow11, 
DBLP:journals/cgf/PothkowWH11,
SchlegelKS12,
poetkow13, poethkow13b, AthawaleE13, 
AthawaleSE16}. More related to our work, several approaches have been 
investigated to estimate the positional variability of critical points under a 
Gaussian PDF model \cite{otto10, OttoGT11, petz12, LiebmannS16} or 
interval-based pointwise representations of the uncertainty \cite{bhatia12, 
szymczak13, tierny_ev14}. However, a common aspect of the above techniques is 
that they explicitly rely on strong assumptions regarding the probability 
density function (PDF) modeling the random variables, which are often assumed 
to be Gaussian or uniform (which is implicitly the case for interval-based 
representations). 
Such strong assumptions are 
 limiting in practice when 
considering ensemble data, where PDF reconstructed from the empirical 
observations can follow an arbitrary, unknown model. Moreover, most of the 
parametric PDF models studied in these techniques do not consider multi-modal 
distributions, which is \brian{however necessary}{a necessity} when several trends clearly occur in 
the ensemble.

\noindent
\textbf{(ii) Ensemble visualization:} A different category of techniques has
been specifically investigated to visualize variability in ensemble data. In 
this setting, a series of global empirical observations (i.e. the members of 
the ensemble) are taken 
as an input for the actual computation of 
\brian{geometrical}{geometric} 
constructions, such as level sets or streamlines\brian{, and t}{. T}he 
\brian{spatial}{}
variability of \brian{these}{the} constructions 
is then analyzed. For instance, 
spaghetti plots \cite{diggle02}
have been used intensively to visualize level set variability in weather 
ensemble data \cite{DBLP:conf/icdm/PotterWBWDPJ09, 
DBLP:journals/tvcg/SanyalZDMAM10}. More advanced representations 
\cite{pfaffelmoser13} have also been proposed, for instance with the notion of 
contour 
boxplot \cite{WhitakerMK13}\brian{ and}{,} their generalization to arbitrary 
curves 
\cite{DBLP:journals/tvcg/MirzargarWK14}, and also 
\brian{applied}{their application} to weather forecast 
data \cite{DBLP:journals/tvcg/QuinanM16}.
\brian{}{Additionally,} Hummel et 
al. derive a complete framework for the visualization of the variability in 
particle advection in ensemble data \cite{DBLP:journals/tvcg/HummelOGJ13}. More 
related to our work, \brian{in particular}{particularly} in \brian{its}{their} ambition to analyze trend 
variability, \brian{specifically-designed}{specialized} clustering techniques have been investigated 
 to cluster isocontours \cite{FerstlKRW16} and streamlines 
\cite{DBLP:journals/tvcg/OeltzeLKJTP14, FerstlBW16}\brian{ to identify and analyze 
trends in the distribution of these geometric constructions within ensemble 
data.}{.} However, \brian{to  our knowledge, }{}this overall strategy has never 
been studied for topological \brian{constructions}{features} such as critical points\brian{, which we do 
in this paper.}{.}
Our approach aims at identifying trends in critical point layouts within 
the ensemble. 
\julien{Therefore}{Thus}, a necessary building block of our framework is a 
method 
to evaluate the similarity between critical point distributions. More 
generally, the similarity estimation between topological data representations is 
a long-studied problem. Several 
\julien{heuristical approaches}{heuristics} have been proposed 
to quickly assess structural similarity \cite{hilaga:sig:2001, 
SaikiaSW14_branch_decomposition_comparison, thomas14}. At the theoretical 
level, several metrics have been carefully studied to evaluate the similarity 
between persistence diagrams \cite{cohen-steiner05, interleaving_distance}, 
merge trees \cite{BeketayevYMWH14} or Reeb graphs \cite{bauer14}. The 
computation of barycenters of such constructions (which is relevant to 
clustering) has also been studied \cite{TurnerMMH14}. However, while 
extensively studied from a theoretical perspective, 
the evaluation of these metrics involve\julien{}{s} computationally 
\julien{intensive}{expensive}
combinatorial 
optimization methods \cite{munkres}\brian{, which}{. This} makes 
them challenging to compute, \brian{if not even }{and potentially }impracticable 
for real-life cases \cite{botnan18}. 
\julien{}{Alternatively, kernel based methods 
\cite{ReininghausHBK15, 
CarriereCO17} have been 
specifically developed for measuring distances between persistence diagrams in 
machine learning tasks.}
\julien{More importantly,}{However,}
these approaches \julien{}{(metric or kernel based)} focus on the intrinsic 
structural similarity between 
topological constructions without considering the \brian{geometrical}{spatial} realization of 
these structures in the original\brian{geometrical}{} domain.

In contrast, 
our approach 
based on persistence maps leverages the soundness of 
topological persistence
\cite{edelsbrunner02} while specifically focusing on the \brian{geometrical}{spatial} layout of 
the critical points.

\subsection{Contributions}
This paper makes the following new contributions:
\begin{enumerate}[leftmargin=1em]
  \item{
  \vspace{-1ex}
  \emph{An informative representation of critical point layouts} 
(\autoref{sec_persistenceMap}): Inspired by distance field transforms, we 
introduce the \emph{Persistence Map}, a measure of the 
\brian{geometrical}{spatial} density \brian{in}{of}
critical points
which leverages the robustness to noise of topological 
persistence to better emphasize salient features.
\brian{C}{In addition, c}omputations \brian{}{to construct 
this map }are shown to be trivially parallelizable.}
Experiments demonstrat\brian{e}{ing} the relevance of this representation for comparing 
critical point distributions with standard density distance measures \brian{.}{ 
are provided.}
  \item{
  \vspace{-1ex}
  \emph{A statistical space for critical point layouts} 
(\autoref{sec_clustering}):
  We present a  framework which leverages spectral embedding to 
  represent persistence maps as points in a low-dimensional Euclidean space, 
  where distances between points represent dissimilarities in critical point 
layout and where
statistical \brian{tasks}{analysis} can be easily carried out. 
The first two dimensions of this space can be used to generate planar views of 
the ensemble
to visualize the distribution of critical point layouts 
at a global level. We additionally show how 
persistence maps can be clustered in this 
space to reveal the major trends in critical point layouts and how 
relevant automatic suggestions for the number of clusters can be estimated.
  }
  \item{
  \vspace{-1ex}
  \emph{Confidence regions for clusters of critical point 
layouts}
(\autoref{sec_confidenceRegions}):
  Based on the clustering of persistence maps, we show how the notion of 
mandatory critical point \cite{tierny_ev14} can be leveraged to visualize in 
the \brian{geometrical}{spatial} domain the possible outcomes in terms of critical point 
layouts, in particular\brian{,}{} with the visualization of 
confidence regions along with 
their respective probability of appearance. The prediction  accuracy of these 
regions is quantitatively evaluated and compared to a baseline 
strategy using an off-the-shelf clustering approach.
  }
  \item{
  \vspace{-1ex}
  \emph{Implementation:} We provide a lightweight C++ 
implementation of our approach that can be used for reproduction purposes.}
\end{enumerate}

\section{Preliminaries}
This section 
\julien{
briefly describes our formal setting. We refer the reader to 
reference books for introductions to Morse theory \cite{milnor63} and 
computational topology \cite{edelsbrunner09}.}{presents 
theoretical background on topological data analysis (TDA). It contains 
definitions adapted from Tierny et al.~\cite{ttk17}. Reference introduction 
books to Morse theory and computational topology have been published by 
Milnor~\cite{milnor63} and Edelsbrunner and Harer~\cite{edelsbrunner09}.}

\subsection{Background}
\label{sec:background}
\julien{
We assume our data is\textbf{Data representation:}}{ The input data is given 
as an ensemble of $n$ piecewise (PL) linear scalar fields on a
PL $d$-manifold $\domain$ (with $d = 2$ or $3$)
$f^{(x)} : \domain \rightarrow \mathbb{R}$, \julien{}{with} $x \in [0, ~n-1]$.} 
Each individual scalar field $f^{(x)}$
is an \emph{ensemble member}. For brevity, 
we will only use the $^{(x)}$ notation in cases where several members 
are considered and thus need disambiguation.
\julien{A member $f$ has value at the set of vertices $\domain^0$ of $\domain$ 
and 
assumed to be linearly interpolated on the simplices of higher dimension.}{Each 
member $f$ is valued at the vertices $\domain^0$ of $\domain$ and is linearly 
interpolated with barycentric coordinates on the remaining simplices of 
$\domain$.}
\julien{We assume and enforce that $f$ is}{In practice, $f$ is enforced to be
injective 
\julien{for the vertices of}{on} $\domain^0$ 
\julien{via a variation of}{with}
simulation of simplicity \cite{edelsbrunner90}.
  The set of simplices 
  \julien{that have}{having}
  a given simplex $\simplex$ as a face form the \emph{star} 
\julien{}{of $\simplex$, } $\stt{\simplex}$.  The set of faces of the 
simplices \julien{in}{of} $\stt{\simplex}$
that do not intersect $\simplex$ form the \emph{link} \julien{}{of $\simplex$,} 
$\lk{\simplex}$.}
\julien{The topology of our domain can be described by its \emph{Betti 
numbers}, where $\beta_i$ are the ranks of its homology 
groups~\cite{edelsbrunner09}.}{}

\brian{}{
  For a vertex $v$, let us define $\lkminus{v}$
to be the \emph{lower link} \julien{}{of $v$} ($\lkminus{v} = \{ \sigma \in 
\lk{v} ~ | ~ 
\forall u \in \sigma : f(u) < f(v)\}$) and $\lkplus{v}$ as the 
\emph{upper link} \julien{}{of $v$} ($\lkplus{v} = \{ \sigma \in \lk{v} ~ | ~ 
\forall u \in \sigma : f(u) > f(v)\}$). 
  When both $\lkminus{v}$ and 
$\lkplus{v}$ are simply connected, $v$ is called a \emph{regular vertex}.  
\julien{Otherwise}{If not}, $v$ is a \emph{critical 
point} of $f$ \cite{banchoff70}.  
\julien{}{Such points correspond to configurations where the \emph{sub-level 
sets} $\sub{i}$ of $f$ (subset of $\domain$ valued below the isovalue $i$)} 
change their topology
when continuously varying the isovalue $i$.
\julien{Given a $\domain$ of dimension $d$, }{
Critical points often correspond to features of interest in applications. 
They are usually classified with the notion of
\emph{index} $\Index$}, which is equal to $0$, $1$, $(d-1)$ and $d$ 
respectively for:
minima (empty lower link),
\julien{}{$1$-saddles (2 connected components of lower link),}
$(d-1)$-saddles (2 connected components of upper link), and
maxima (empty upper link) respectively.
}

\begin{figure}
  \includegraphics[height=2.15cm]{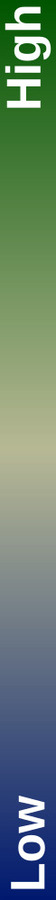}
  \hfill
 \includegraphics[height=2.15cm]{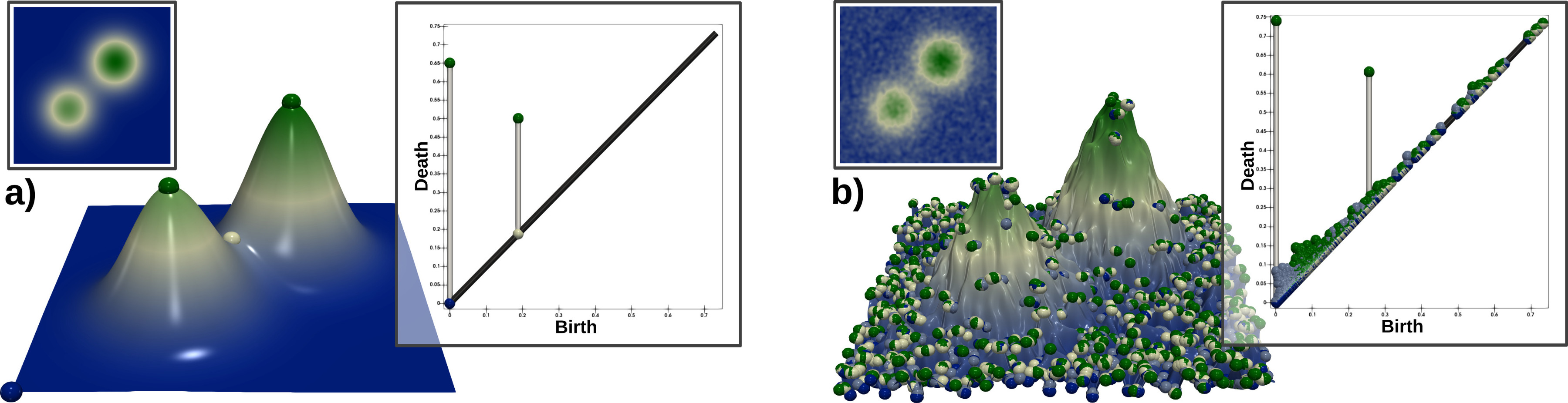}
 \vcaption{
 Critical points (spheres, dark blue: minima, dark green: maxima, other: 
saddles) and persistence diagrams of a clean (a) and noisy (b) 2D scalar 
field. 
From left to right: original 2D 
data, 3D terrain 
representation, persistence diagram. The diagrams clearly exhibit in both cases 
two large pairs, corresponding to the main hills. In the noisy diagram 
(b), small bars near the diagonal correspond to noisy features in the 
data.}
  \label{fig_persistenceDiagram}
\end{figure}

\label{sec_persistence}
\brian{}{
The \julien{}{population of} critical points of $f$ can be 
\julien{illustrated}{visually encoded} 
\julien{by a topological abstraction 
called the}{with the notion of} \emph{persistence 
diagram}~\cite{edelsbrunner02} (\autoref{fig_persistenceDiagram}). This diagram
encodes critical points as pairs \julien{such that}{}$(c, 
c')$ \julien{with}{such that} $f(c) < f(c')$ 
and $\Index(c) = \Index(c') - 1$.  
These pairs follow the Elder rule~\cite{edelsbrunner09},
\julien{such}{which intuitively implies}
that if two \julien{}{topological} features of $\sub{i}$ meet at a critical 
point $c'$ of $f$, 
the \emph{youngest} feature
(created at the highest function value)
 \emph{dies}, favoring the \emph{oldest} (created at the lowest function 
value). 
In a persistence diagram $\persistenceDiagram{f}$,
each pair  $(c, c')$ is represented as a point in 2D at coordinates 
$\big(f(c), f(c')\big)$, which are
the  \emph{birth} and \emph{death} of the pair respectively.  
The  \emph{persistence} of the pair
is given by its height in the diagram, $P(c, c') = |f(c') - f(c)|$. It
describes the lifespan \julien{}{in the range} of \julien{a}{the corresponding} 
topological feature.} 
\julien{
In the rest of the paper, when discussing persistence diagrams, 
we will only 
consider critical point pairs of index $(0, 1)$ and $\big((d - 1), d\big)$.}{In 
the following, only the critical point pairs involving local extrema,  $(0, 1)$ 
and $\big((d - 1), d\big)$, will be considered.}
\julien{The 
impact of this simplification is discussed in \autoref{sec_results}.}{The 
consequence of this simplifying assumption are described in 
\autoref{sec_results}.}
\julien{}{Moreover, for genericity purposes, 
all persistence evaluations will be normalized with regard to the largest 
persistence found in the data ($P(c, c') \in [0, 1]$).}
\julien{}{In practice, 
the pairs of the diagram located in the vicinity of the diagonal denote 
low-amplitude noise
while prominent features will be associated with persistent pairs, located
far away from the diagonal (\autoref{fig_persistenceDiagram}).}
The 
persistence diagram has
been extensively studied from a theoretical perspective and its stability
to perturbations in the input data has been demonstrated 
\cite{cohen-steiner05}. This
stability result greatly motivated the use of persistence in applications,
ranging from machine learning \cite{chazal13} to visualization, where it has 
been shown to significantly help users distinguish salient features from noise.

\begin{figure*}
  \includegraphics[height=3.95cm]{colorMap_blueGreen.jpg}
  \hfill
  \includegraphics[height=3.95cm]{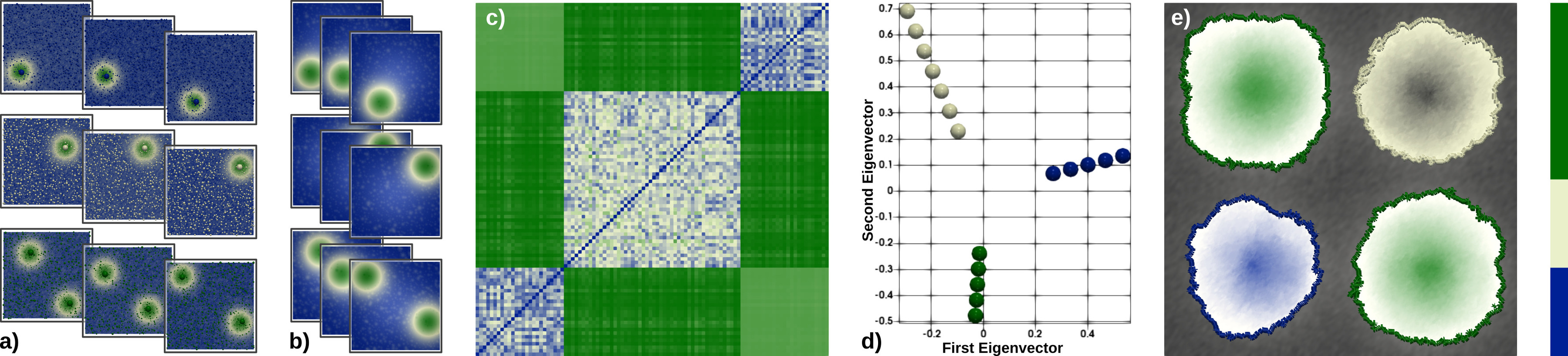}
 \vcaption{
 Pipeline overview. 
 (a) Given an ensemble of scalar fields, our approach 
computes the persistence diagram of each member (critical points are scaled by 
persistence). (b) To ease the comparison of critical point layouts, each member 
is transformed into an alternative representation, the persistence map, which 
denotes the local density in salient features. (c) The distance matrix 
between the persistence maps is constructed and used to embed each member 
in a low dimensional \julien{}{feature} space, whose
first two components can be used to generate planar overviews of the distinct 
critical point patterns found in the ensemble (d). In this \julien{}{feature} 
space, clusters of 
members are automatically estimated. (e) The persistence atlas is 
finally composed by considering the mandatory critical points of each cluster 
independently (one color per cluster), revealing the main configurations of 
critical point layouts in the ensemble in terms of numbers and positions. 
\julien{The likelihood of appearance of critical points}{The positional 
variability of critical points}
within each 
mandatory critical point is \julien{shown}{indicated} with a color map.
The 
bar plot (right side) \julien{indicates}{shows} the statistics of appearance of 
each 
cluster in the ensemble.
%
%
}
 \label{fig_overview}
\end{figure*}

\subsection{Overview}
Our approach is composed of three main steps (\autoref{fig_overview}). It takes 
as \brian{an }{}input $n$ PL scalar fields defined on the same PL manifold 
$\domain$.

First (\autoref{sec_persistenceMap}), the persistence map of each ensemble 
member is computed. The purpose of this representation is to evaluate the 
spatial distribution of the critical points in each member, while 
\brian{}{at the same time }balancing the 
contribution of each critical point by its 
persistence\brian{, in order}{} to 
emphasize salient features and reduce the contribution of noise. 
\BrianSays{don't like balace in prev sentence}

Second (\autoref{sec_clustering}), we leverage spectral embedding to represent 
each member as a point in a low-dimensional Euclidean space\brian{, where d}{. D}istances in 
this feature space denote dissimilarities between persistence maps. 
This space is conducive to further statistical analysis of the members which 
are clustered based on their persistence maps.
The first two dimensions of this space are used to generate planar views
which
enable the direct visualization of
the main trends in the ensemble in terms of critical point layouts. 

Third (\autoref{sec_confidenceRegions}), confidence regions in the geometrical 
domain are computed for each cluster by leveraging the notion of mandatory 
critical point\brian{}{s} \cite{tierny_ev14}. Finally, the confidence regions of all 
clusters are composed together into the final persistence 
atlas\brian{, which}{. This} enables 
the visualization of the regions of occurrence of the most prominent critical 
points along with estimations of their probability of appearance.

\section{Persistence maps}
\label{sec_persistenceMap}
In this section, we introduce the notion of persistence map, a representation 
of the critical point distribution in each member. 

\subsection{Motivation}
The main target of 
persistence maps is to facilitate the comparison of two members $f^{(x)}$ 
and $f^{(y)}$ in terms of the layout of their critical points.
As discussed in \autoref{sec_relatedWork}, existing topological metrics (e.g. 
the Bottleneck distance \cite{cohen-steiner05}) do not take into account the 
spatial embedding of the critical points in $\domain$ and are therefore not 
suited for our purpose.
Let $C^{(x)}$ and $C^{(y)}$ be the set of critical points of $f^{(x)}$ and 
$f^{(y)}$ respectively, which can be interpreted as point clouds in $\domain$. 
The problem of comparing the layouts of critical points of $f^{(x)}$ 
and $f^{(y)}$ then reduces to that of comparing two point clouds, a problem for 
which no universal solution exists. For instance, the Haussdorff 
distance, \brian{which }{}can be seen as a \emph{worst-case} \brian{metric, 
}{metric that} only measures the distance 
between the two most distant points of the two sets\brian{, which}{.  This} is too limiting for 
our setting since the similarity of the rest of the point cloud is 
not assessed.
Moreover, due to the presence of noise in the data, it is highly likely in 
practice that a significant number of the critical  points of $f^{(x)}$ and 
$f^{(y)}$ are noise artifacts\brian{, even in the case of smooth simulation data}{}.
Such \brian{noise }{}artifacts must be taken into account in the similarity estimation
in order to reduce \brian{the}{their} importance\brian{ of noise}{} and highlight salient features. This 
last observation is the main motivation behind persistence maps.

A reason for the difficulty in estimating the similarity between the 
point clouds  $C^{(x)}$ and $C^{(y)}$ is that  
there exists no canonical parameterization of these sets allowing for a
straightforward comparison with established distance measures, as 
can be done for streamlines for instance \cite{FerstlBW16}. 
 $C^{(x)}$ and $C^{(y)}$  may not even be of the same size.
This observation 
motivates the transformation of $C^{(x)}$ and $C^{(y)}$ into an alternate 
representation that would yield a natural parameterization directly usable with 
standard distance measures. 

\begin{figure}
  \includegraphics[height=3.85cm]{colorMap_blueGreen.jpg}
  \hfill
  \includegraphics[height=3.85cm]{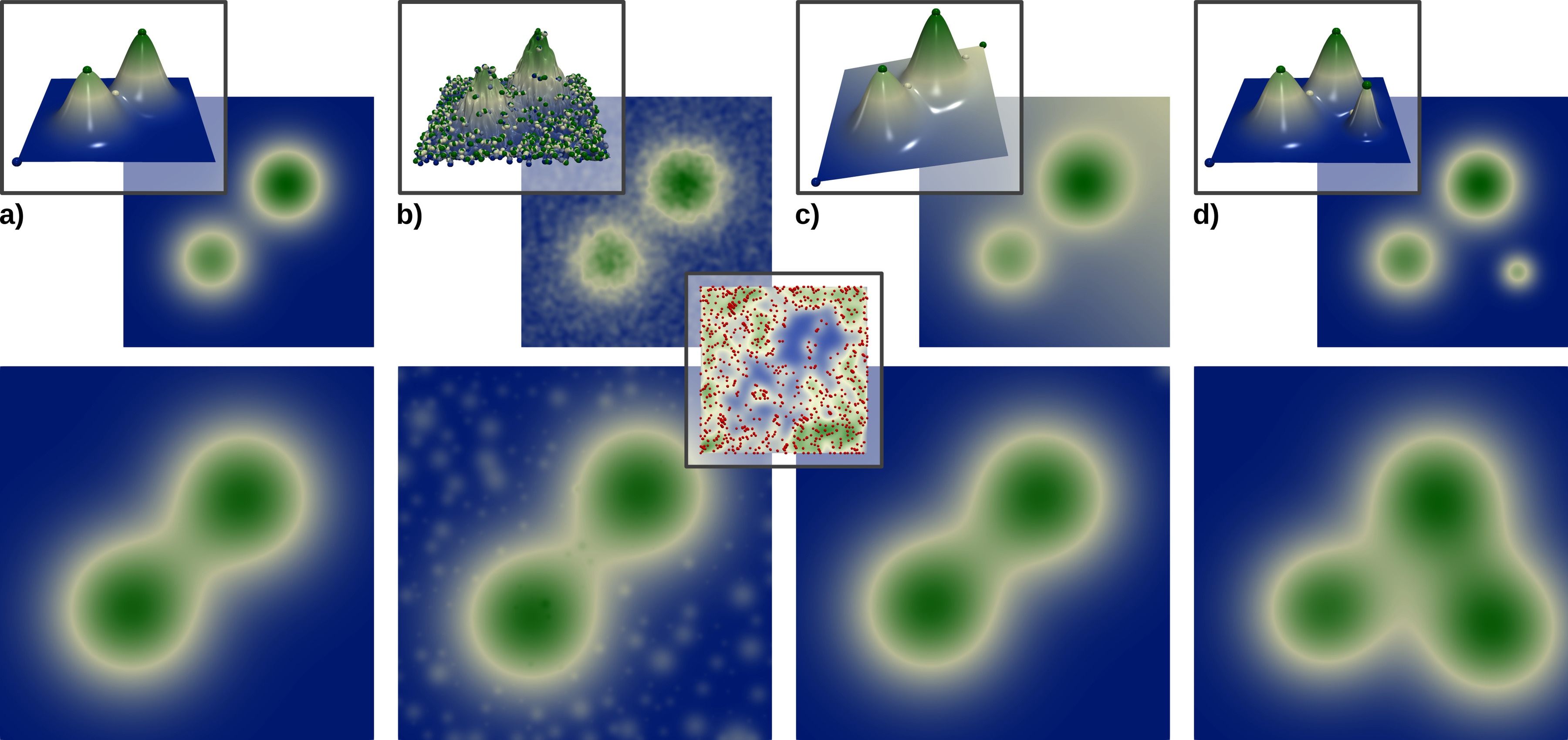}
 \vcaption{
 Persistence maps (bottom, maxima only) for four 2D scalar fields (top, inset: 
terrain view with critical points): $f^{(0)}$ (a), $f^{(1)}$ (b), $f^{(2)}$ 
(c), $f^{(3)}$ (d).
  When using constant \julien{functions}{values} for $\alpha(c)$ and 
$\sigma(c)$, $\phi$ 
estimates the local density in critical points (b, right inset with maxima in 
red). In this example, the $L_2$ distance separating $f^{(0)}$ from $f^{(1)}$, 
$f^{(2)}$ and $f^{(3)}$ is 
$200.79$, 
$41.03$ and 
$14.18$ respectively. In 
contrast, the same distance from $\phi^{(0)}$ to $\phi^{(1)}$, 
$\phi^{(2)}$ and $\phi^{(3)}$ is 
$11.09$, 
$2.84$ and 
$49.01$ respectively. 
This different ordering indicates that the $L_2$ distance between persistence 
maps is less sensitive to noise and global shifts in data values and better 
discriminates changes in salient features. 
 }
 \label{fig_persistenceMap}
\end{figure}

\newcommand{\eqspace}{\vspace{-1ex}}

\subsection{Formulation}
\label{sec_persistenceMap_eq}
Breckner and M\"oller \cite{DBLP:journals/cgf/BrucknerM10} faced a similar 
problem in the context of isosurface comparison and introduced a signed 
distance field transform, measuring the distance between each vertex of 
$\domain$ and the considered isosurface. 
Then, the similarity between two isosurfaces can be evaluated based on the 
standard $L_2$ distance between their distance transforms.
The same idea has been later used by 
Ferstl et al. \cite{FerstlKRW16} in a context that\brian{highly}{} resembles 
our 
setting
(isocontour clustering for level set variability analysis and visualization). 
We build upon this strategy \brian{for}{to construct} persistence maps. In particular, one could 
derive \brian{such}{} a distance transform for a critical point set $C$, by 
considering for each vertex $v \in \domain$, the distance to the closest 
critical point of $C$. 
However, such a distance transform would be highly \brian{sensible}{sensitive} to the presence of 
noise in the data since all the critical points of $f$ would be 
considered for its computation. 
Therefore, it 
is necessary to develop a 
transformation where the contribution of each critical point could be weighted 
by an importance measure, such as topological persistence 
\cite{edelsbrunner02}. While such a weighting strategy is difficult to elaborate 
for distance fields, it is much easier to derive for sums of gaussian radial 
basis functions. In particular, let $\phi : \domain \rightarrow \mathbb{R}^+$ 
be the following scalar function, where $\alpha(c)$ and $\sigma(c)$ are scalars
controlling the 
amplitude and spatial spread of the contribution of \julien{}{the critical 
point} $c$:
\begin{eqnarray}
\eqspace
  \label{eq_persistenceMap}
 \phi(v) = \sum_{c \in C} \alpha(c) e^{-{{||v - 
c||_2^2}\over{2 \sigma(c)^2}}}
\eqspace
\end{eqnarray}
If constant values are considered for both $\alpha(c)$ and $\sigma(c)$, $\phi$ 
is a measure of the local critical point density 
(\autoref{fig_persistenceMap}, inset).
To limit the importance of 
noisy critical 
points in this density estimation and to highlight  salient 
features\julien{instead}{}, we 
use\julien{topological}{} persistence as an importance measure in the 
expressions of  
$\alpha(c)$ and $\sigma(c)$  as follows, where $P(c)$ stands for the persistence 
of the critical point pair containing $c$ in $\persistenceDiagram{f}$:
\begin{eqnarray}
\eqspace
 \alpha(c) =  P(c), \quad \sigma(c) = \gamma P(c)
\eqspace
 \label{eq_persistence}
\end{eqnarray}
$\gamma$ 
controls the focus 
that is
given 
to salient features in terms of their spread in the \brian{geometrical}{spatial}
domain.
\brian{(d}{D}istances are normalized with regard to the bounding box 
diagonal\brian{)}{}. We have found that 
$\gamma = 0.1$ 
is a good value in practice. 
\julien{Note 
that 
t}This representation resembles the 
notion of persistence images \cite{persistenceImages}, which 
focuses on range rather than domain density.


\subsection{Distances}
\label{sec_distanceMatrix}
Since they are both defined on the same \brian{geometrical}{spatial} domain $\domain$, the 
persistence maps $\phi^{(x)}$ and $\phi^{(y)}$ of two critical points sets 
$C^{(x)}$ and $C^{(y)}$ benefit from a common parameterization and their 
distance can be estimated with standard distance measures, such as the $L_2$ 
norm:
\begin{eqnarray}
\eqspace
 ||\phi^{(x)} - \phi^{(y)}||_2 = \sqrt{\sum_{v \in \domain} \big(\phi^{(x)}(v) 
- \phi^{(y)}(v)\big)^2}
\eqspace
\label{eq_distance}
\end{eqnarray}
By design, this metric is robust to noise, since the contribution of critical 
points to the persistence maps is weighted by their 
persistence \julien{.}{ (\equref{eq_persistence}). Hence, small persistence 
pairs 
(typically corresponding to low amplitude noise, 
\autoref{fig_persistenceDiagram}(b))
will have a 
negligible contribution in practice to the persistence maps 
(\autoref{fig_persistenceMap}(b),
further discussion in 
\autoref{sec_limitations}).
} This is 
important since small scale additive noise often occur in practice even for 
\brian{}{assumed} smooth simulation data. 
This metric is also robust by design to global 
variations in data values which do not change the critical point spatial 
layout, since the actual data values are not taken into account in the 
persistence map. In contrast, 
the standard $L_2$ distance $||f^{(x)} - f^{(y)}||_2$
would tend to miss the possible preservation 
of salient features in the presence of global shifts in data values,
as\brian{it}{} can 
be the case with seasonal effects in climate data. 
Finally, the distance 
$||\phi^{(x)} - \phi^{(y)}||_2$ 
is specifically designed to penalize changes in the layout of salient 
critical points.
The above properties are illustrated in \autoref{fig_persistenceMap}, which 
shows persistence maps on a toy example, 
$f^{(0)}$, along with three variants: $f^{(1)}$ with additive noise, $f^{(2)}$ 
which contains a global shift in data values (slope), and $f^{(3)}$ which 
contains an additional salient feature. For this data, we have:
$||f^{(0)} - f^{(3)}||_2 < ||f^{(0)} - f^{(2)}||_2 < ||f^{(0)} - f^{(1)}||_2$. 
In other words, with the $L_2$ distance between the actual data values, 
the noise affected dataset ($f^{(1)}$) is the most distant to the original \brian{one}{}
($f^{(0)}$), while the dataset with a drastic change in critical point layout 
($f^{(3)}$) is the closest. In contrast, the $L_2$ distance between the 
corresponding persistence maps results in a different ordering:
$||\phi^{(0)} - \phi^{(2)}||_2 < ||\phi^{(0)} - \phi^{(1)}||_2 < ||\phi^{(0)} - 
\phi^{(3)}||_2$. In other words, with the persistence map metric, the closest 
data set from 
the original \brian{one}{} ($f^{(0)}$) is the one which better preserves the critical 
point layout ($f^{(2)}$), while the most distant \brian{one}{ }is the one which 
changes it the most ($f^{(3)}$). 
This indicates that the metric $||\phi^{(x)} - \phi^{(y)}||_2$ 
is indeed more robust to noise and global shift in data values than 
$||f^{(x)} - f^{(y)}||_2$ and that it better describes variations in 
the layout of salient critical points.
\julien{}{Our distance (Eq.~\ref{eq_distance}) resembles the 
\emph{kernel distance} defined for generic point cloud data 
\cite{PhillipsWZ15}. In contrast, persistence maps focus on the critical points 
of a scalar field (instead of generic point clouds). This allows to 
additionally consider in the density estimation 
the persistence of each critical point 
 as an importance measure
(Eq.~\ref{eq_persistence}), 
to highlight salient 
features and reduce the effect of noise.
}


\section{Space of persistence maps}
\label{sec_clustering}

As described above, the $L_2$ distance between persistence maps is a good 
candidate to compare the spatial layout of critical points between two members.
Based on this metric, a distance matrix $\Phi$ is computed
for the entire ensemble, with $\Phi_{xy} = ||\phi^{(x)} - \phi^{(y)}||_2$, 
and then normalized.
In this section, we exploit this distance matrix to visualize and 
identify the main trends in critical point layouts within the 
ensemble.


\subsection{Low dimensional embedding}
\label{sec_eigenMap}
To directly visualize the global trends in critical point layouts, we first 
consider a low dimensional embedding of the ensemble into a space of 
persistence maps, noted $\mathcal{P}$,
where each map $\phi$ is represented by a point and where distances between 
points denote distances between persistence maps. For this, we employ 
established 
methods
for non-linear dimensionality 
reduction \cite{wickelmaier2003introduction, borg05}. 
In particular, we focus on the spectral approach by Belkin et al. 
\cite{belkin2003laplacian} based on Laplacian eigenmaps, which has been shown 
to 
better preserve locality than standard methods such as principal component 
analysis \cite{surveyPCA} or Isomap \cite{tenenbaum00}. This property is 
particularly beneficial if clustering is subsequently considered, which is the 
case in our framework (\autoref{sec_subClustering}). For completeness, we 
briefly sketch the main steps of the Laplacian eigenmap approach and we refer 
the reader to 
\cite{belkin2003laplacian} for further details.

First, an adjacency graph is constructed, where the $x^{th}$ node represents
the $x^{th}$ ensemble member and where arcs are introduced between the node 
$x$ and its $n_n$ nearest neighbors 
(according to the distance matrix $\Phi$). 
In practice, we set $n_n$ to a default recommended value ($5$).
 Next, a weight matrix $W$ is constructed such that
$
W_{xy} = 1
$
if $x$ and $y$ are connected in the adjacency graph and $0$ otherwise.
A diagonal matrix $D$ is also established such that $D_{xx} = \sum_{y} 
W_{xy}$. Then, the Laplacian, $L$, of the adjacency graph is considered as $L = D 
- W$, which is a symmetric, positive semidefinite $n \times n$ matrix 
\cite{belkin2003laplacian}. Finally, 
the low-dimensonal space $\mathcal{P}$ is constructed by 
projecting each ensemble member along the $n_d$ first eigenvectors 
$\psi \in \mathbb{R}^n \times \mathbb{R}^n$ of $L$, which are solutions of the 
generalized eigenvector problem: $L \psi = \lambda D \psi$ (where $\lambda \in 
\mathbb{R}^n$ stands for the $n$ eigenvalues of $L$). In practice, the first 
eigenvector $\psi_0$ is discarded, as suggested by Belkin et al. 
\cite{belkin2003laplacian}. Thus, the $x^{th}$ ensemble member is then 
embedded at position 
$\psi^{(x)} = \big(\psi_1(x), \dots, \psi_{n_d}(x)\big)$.
Since the first eigenvectors of $L$ are usually considered to be the most 
informative \cite{vonLuxburg2007}, for visualization purpose, we typically 
represent planar layouts of the space of persistence maps $\mathcal{P}$ by only 
considering the first two components of this vector ($\psi_1(x), \psi_2(x)$).

%


\subsection{Persistence map clusters}
\label{sec_subClustering}

\autoref{fig_overview}(d) shows a typical 2D layout of the first two dimensions 
of 
 the space of persistence maps, $\mathcal{P}$, 
for a 
toy ensemble dataset. As shown in this example, clear \brian{patterns,}{patterns 
that}
correspond\brian{ing}{} to distinct trends in critical point layout emerge from this 
visualization. To quantitatively analyze these patterns, we next employ 
 clustering algorithms.
In 
particular, we employ the popular $k$-means algorithm \cite{historicKmeans}, 
which has been shown to be well suited for a combined usage with spectral 
emdedding (\autoref{sec_eigenMap}), yiedling the notion of \emph{spectral 
clustering} \cite{DBLP:conf/cvpr/ShiM97}. This algorithm is based on the 
classical Lloyd relaxation scheme \cite{lloyd57} which,
given an initial assignment of $k$ cluster centroids chosen among the data 
points, 
assigns each data point to the cluster of its closest 
centroid. Next, for each cluster, a new centroid is selected as the point being 
the closest to the new cluster barycenter and the procedure is iterated until 
convergence. Note that for the above clustering procedure, the spectral 
clustering literature recommends to only use the $k$ first components of 
$\psi^{(x)}$ \cite{vonLuxburg2007}, although we found in practice that with our 
implementation, the most stable results were obtained for $n_d = k - 1$.

\begin{figure}
  \centering
  \includegraphics[width=0.9\linewidth]{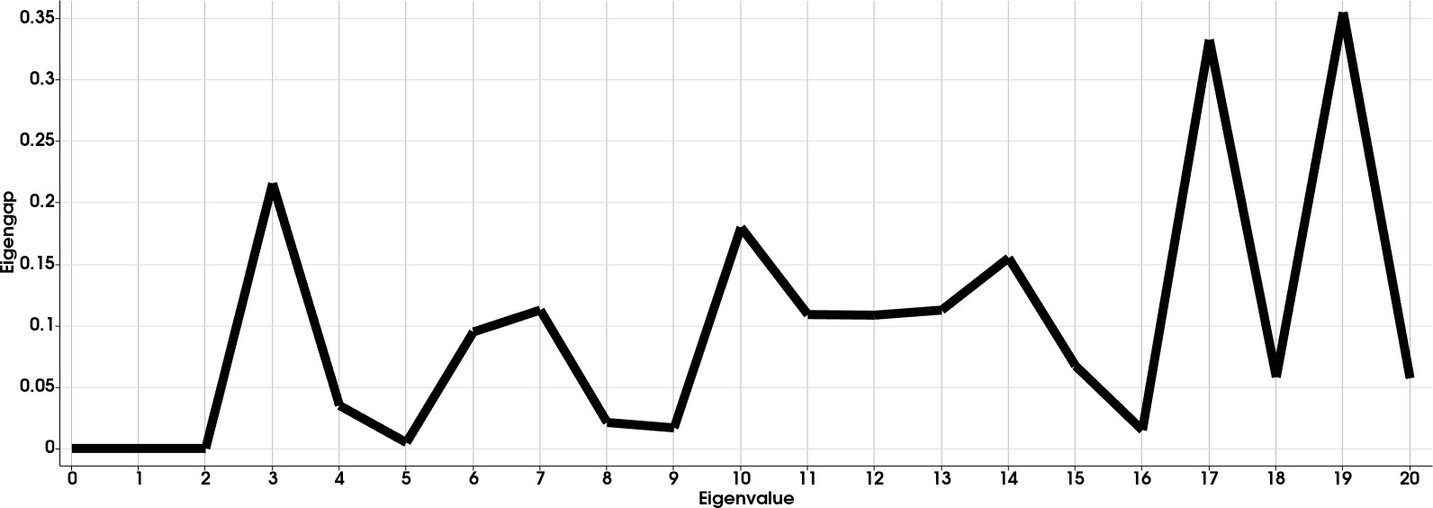}
  \vspace{-1.5ex}
 \vcaption{Eigengaps computed for the example shown in \autoref{fig_overview}. 
Our approach uses the position of the first local maximum of eigengap as an 
initial value for the number of clusters $k$. Our framework also offers the 
possibility to explore the other maxima, as well as arbitrary $k$ values.}
\vspace{-1ex}
  \label{fig_eigenGaps}
\end{figure}

The number $k$ of clusters to be considered is particularly important as it 
directly corresponds to the number of trends which can be visualized in the 
ensemble. While we offer users the possibility to explicitly specify 
pre-defined values of $k$, we also provide an automatic estimation procedure. 
Several statistical measures have been studied for the automatic estimation of 
$k$, such as the Bayesian Information Criterion \cite{pelleg00}. In the 
specific case of spectral clustering however, it has been shown that the 
eigenvalues of the Laplacian matrix (\autoref{sec_eigenMap}) 
already exhibit important hints regarding cluster numbers and
that they are particularly 
useful to identify 
proper values for $k$.
In particular, the first eigenvalue $\lambda_k$ resulting in a significant 
\emph{eigengap} $\delta_k = |\lambda_k - \lambda_{k+1}|$ is usually considered 
as a good value for $k$ (see von Luxburg \cite{vonLuxburg2007} 
for formal 
arguments based on perturbation theory). Thus, in practice, we provide 
as an 
initial guess for $k$, the position of the first local maximum of eigengap 
$\delta_k$.
\autoref{fig_eigenGaps} plots the evolution of the eigengaps for the example of 
\autoref{fig_overview}. As shown in this figure, the appropriate number of 
clusters for this specific dataset indeed corresponds to the first local 
maximum of
eigengap ($k = 3$). Note that several other local maxima of eigengaps occur for 
higher eigenvalues. We also offer users the possibility to interactively 
explore them individually.

%
%
%
%
%
%
%
%

\section{Confidence regions for persistence map clusters}
\label{sec_confidenceRegions}
The major trends in critical point layout in the ensemble 
can be identified by clustering the persistence maps (\autoref{sec_clustering}). 
In this section, we describe how to visualize\brian{, back on the geometrical domain,}{} 
the spatial variability of critical points within each of the identified 
clusters.

\subsection{Per cluster variability analysis}
\label{sec_mandatoryCriticalPoints}
The clustering procedure described in the previous section 
identifies
disjoints subsets of ensemble members which share a common 
pattern in critical point layout. Let $f^{(X)} = \{f^{(x_0)}, f^{(x_1)}, \dots 
f^{(x_m)}\}$ be such a subset ($m < n$). To understand the variability 
of critical points within this subset, one needs  first \emph{(i)} to
identify a common topological structure among all of the members of $f^{(X)}$ 
and second \emph{(ii)} to analyze its spatial variability. 
As discussed in \autoref{sec_relatedWork}, several approaches have been 
proposed to study the positional uncertainty of critical points. Among those, 
we focus on the approach based on mandatory critical points 
\cite{tierny_ev14} 
since it
is
based on point-wise intervals
and is\brian{}{,} therefore\brian{}{,} well suited for the analysis of 
ensemble data, where no specific assumption can be made about the structure of 
the point-wise random variables locally modeling the data variability. For 
completeness, we briefly sketch the main steps of this method and refer the 
reader to \cite{tierny_ev14} for further details.

\begin{figure}
\vspace{-2ex}
  \centering
\includegraphics[width=.9\linewidth]{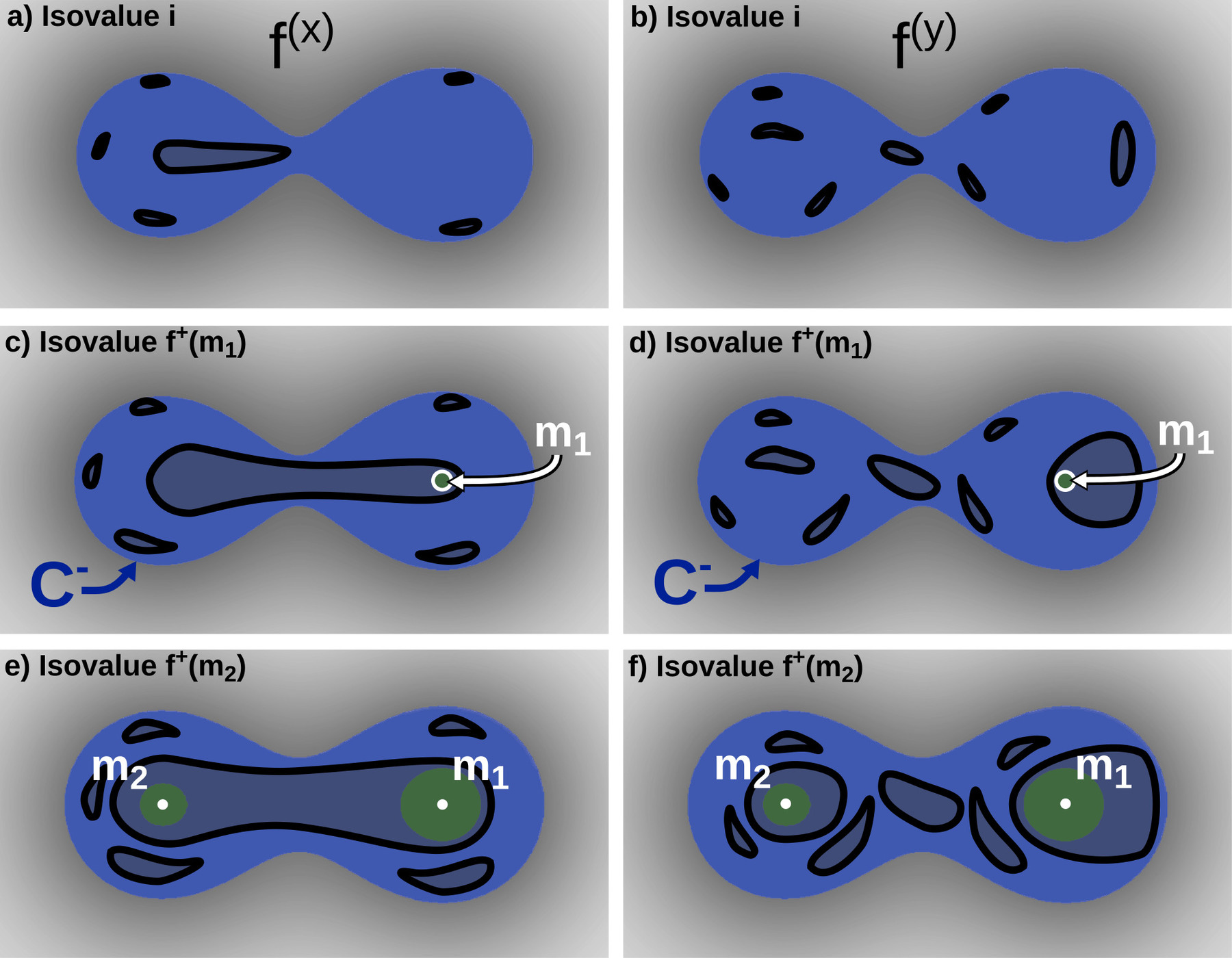}
\includegraphics[width=0.975\linewidth]{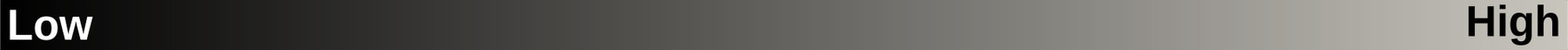}
 \vcaption{
Sub-level sets of $f^-$ (blue) and $f^+$ (green) at three different isovalues 
$i < f^+(m_1) < f^+(m_2)$ with $m_1$ and $m_2$ being minima of $f^+$. The 
sub-level set components $C^{(x)}$ and $C^{(y)}$ of two ensemble members 
$f^{(x)}$ (left) and $f^{(y)}$ (right) are shown in gray.}
 \label{fig_candidateRegions}
\end{figure}

First, pointwise scalar value bounds are extracted as two scalar fields $f^- : 
\domain \rightarrow \mathbb{R}$ and $f^+ : \domain \rightarrow \mathbb{R}$, 
such that $f^-(v) = \min_{f^{(x)} \in f^{(X)}} f^{(x)}(v)$ and $f^+(v) = 
\max_{f^{(x)} \in f^{(X)}} f^{(x)}(v)$. 
Given an isovalue $i$, let $C^-(i)$ be a connected component of sub-level set 
of $f^-$ (blue region in \autoref{fig_candidateRegions}). By construction, for 
each vertex 
$v$ in $C^-(i)$, there exists at least one member $f^{(x)} 
\in f^{(X)}$ for which $f^{(x)}(v) \leq i$. Then, there exists a  
member $f^{(x)}$ 
for which 
a connected component of sub-level set $C^{(x)}$  passes through $v$ at isovalue 
$i$ (gray regions in \autoref{fig_candidateRegions}). Then, $C^-(i)$ is called a 
\emph{candidate region} for the appearance of a 
local 
minimum (responsible for the creation of the component $C^{(x)}$ in $f^{(x)}$).

Let $m_1$ be a minimum of 
$f^+$. Since $f^-$ and $f^+$ are nested, $m_1$ must be located inside a 
connected component of sub-level set of $f^-$ at isovalue $f^+(m_1)$. 
Let $C^-\big(f^+(m_1)\big)$ be that region and let us first consider that $m_1$ 
is the only minimum of $f^+$ in it.
At 
isovalue $f^+(m_1)$, by construction, \emph{all} the members $f^{(x)} \in 
f^{(X)}$ are such that $f^{(x)}(m_1) \leq f^+(m_1)$. This means, that for 
\emph{all} the members $f^{(x)}$ of the subset $f^{(X)}$, there exists a 
connected component of sub-level set 
$C^{(x)}$
passing through $m_1$ (gray components containing $m_1$ in 
\autoref{fig_candidateRegions}(c) and \autoref{fig_candidateRegions}(d)). In 
particular, this connected component was created at an earlier isovalue, at  
one of the vertices of the corresponding candidate region, 
$C^-\big(f^+(m_1)\big)$. Overall, this means that $C^-\big(f^+(m_1)\big)$ must 
contain at least one minimum (responsible for the 
initial creation of the component $C^{(x)}$) for all the members 
of $f^{(X)}$. Thus, the region $C^-\big(f^+(m_1)\big)$ is 
called  a \emph{mandatory minimum}: a minimal connected component $C^-$ 
of $\domain$, associated with a minimal interval $I^- = [min_{v \in C^-} 
f^-(v), min_{v \in C^-}f^+(v)]$, such that any $f^{(x)}$ contains at least one 
minimum $m_1$ in $C^-$ with $f^{(x)}(m_1) \in I^-$.

\autoref{fig_candidateRegions} illustrates this process where candidate regions 
(blue) may contain several connected components $C^{(x)}$ (gray) of 
sub-level set of ensemble members ($m_1$ is shown in green). 
Note that if the candidate region contains a second minimum $m_2$ such that 
$f^+(m_1) < f^+(m_2)$, this implies that the sub-level set of all members pass 
through $m_2$ as well. However, they may do so with the component $C^{(x)}$ 
which already contains $m_1$ (\autoref{fig_candidateRegions}(e)). Thus, the 
existence of such a second minimum $m_2$ does not necessarily imply the 
existence of an additional minimum in $f^{(x)}$, as it is the case in 
\autoref{fig_candidateRegions}(e) (as opposed to 
\autoref{fig_candidateRegions}(f)). As discussed in \autoref{sec_results}, this 
observation may have important practical implications, 
as it may prevent the detection of a mandatory critical point in case of 
high pointwise value variability $|f^+(v) - f^-(v)|$.

Other types of 
mandatory critical points are extracted similarly, as 
described in \cite{tierny_ev14}. Eventually, each cluster $f^{(X)}$ is 
associated with a collection of mandatory critical points, which describe the 
spatial variability of the common topological structure found among its 
members.



\subsection{Global visualization}
The mandatory critical points can be visualized for each cluster independently, 
by displaying each critical component with a colored region. 
Additionally, \julien{a 
likelihood of appearance of critical points can be computed and visualized 
within each region.}{the positional variability of critical points within each 
region can be estimated and visualized as follows.}
Given a histogram 
representation of the data values taken by a vertex $v$ in $f^{(X)}$, we 
estimate this \julien{likelihood}{variability} as the probability of $v$ to 
admit a scalar value 
within the critical interval of each mandatory critical point. Finally, we 
estimate the overall probability of appearance of a mandatory critical point as 
the proportion between the size of $f^{(X)}$ and the total number\brian{}{,} $n$\brian{}{,} of 
members in the ensemble. As shown in \autoref{fig_overview} (right), this 
probability can be visualized in the form of a barplot.
\brian{Overall t}{T}he Persistence Atlas is then \brian{composed of }{created 
from }a collection of confidence maps 
(\brian{which can be }{}composed together) \brian{which provides,}{that provide} for each major trend found in 
the ensemble, confidence regions for the \brian{appareance}{appearance} of critical points\brian{,}{} along 
with their probability of appearance, as well as\julien{an estimation of}{,} 
their \julien{}{individual critical point} spatial 
variability given by the above\julien{likelihood}{} 
estimation \julien{\brian{,}{.}
\brian{as shown in}{See }
\autoref{fig_overview}.}{(\autoref{fig_overview}).}

\section{Results}
\label{sec_results}

This section presents experimental results obtained on a desktop computer with 
a Xeon CPU (2.6 GHz, 2x6 cores), with 64 GB of RAM. For the computation 
of the persistence diagrams, we used the 
Topology ToolKit (TTK) \cite{ttk17}. 
For the spectral embedding and clustering, we adapted  classes 
from the \emph{scikit-learn} package \cite{scikitlearn}. The other components 
of our 
approach have been implemented as TTK modules.

\begin{figure}
\centering
  \vspace{-2ex}
%
%
\includegraphics[height=5cm]{colorMap_blackWhite.jpg}
\hfill
\includegraphics[height=5cm]{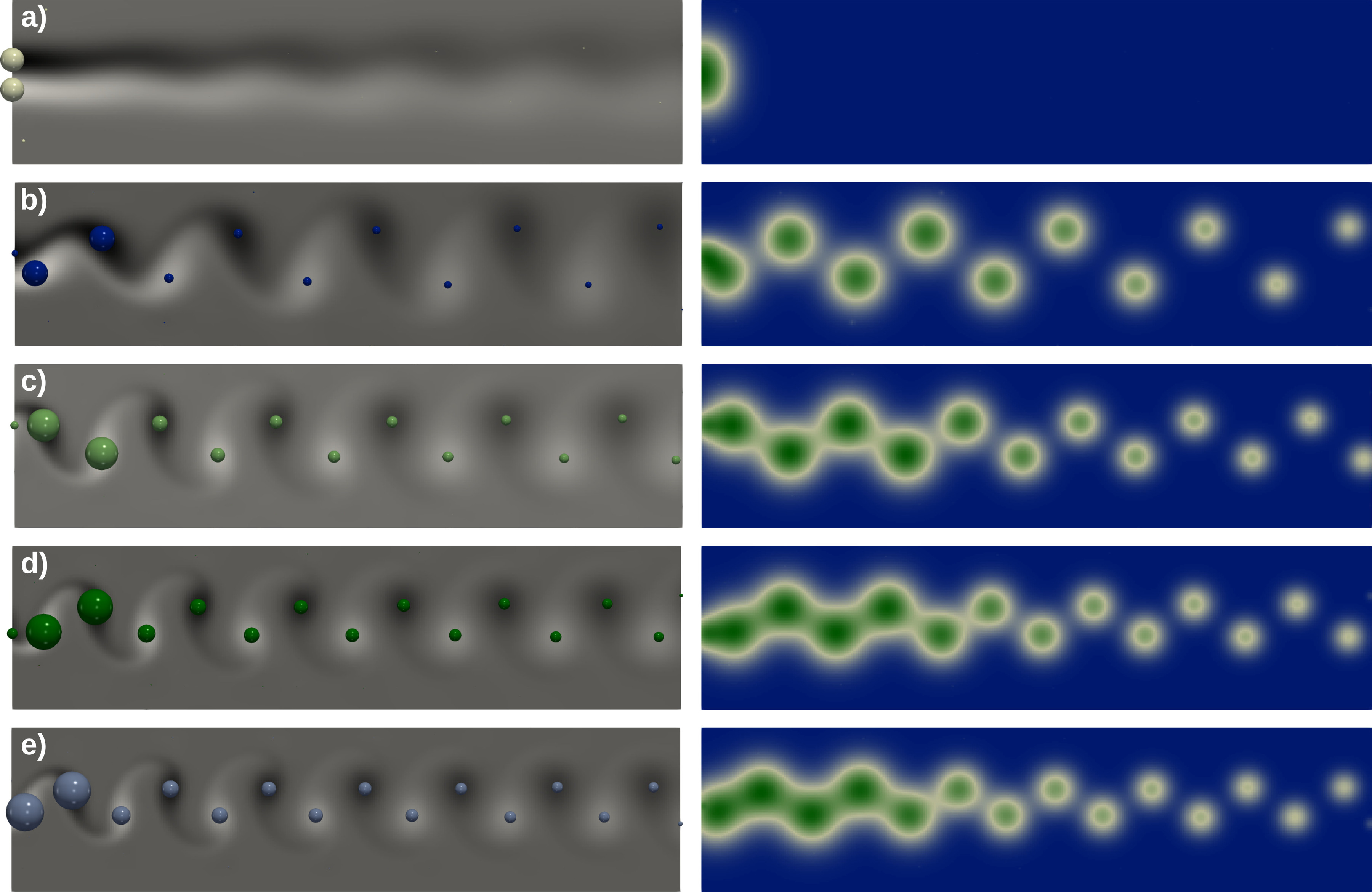}
\hfill 
\includegraphics[height=5cm]{colorMap_blueGreen.jpg}
 \vcaption{Five representative members of the ensemble of
\autoref{fig_teaser}. Critical points (minima and 
maxima) are shown with spheres scaled by their persistence (left). 
Persistence maps are shown on the right.}
  \label{fig_vortexStreet_pm}
\end{figure}

\begin{figure}
  \vspace{-2ex}
 \includegraphics[width=\linewidth]{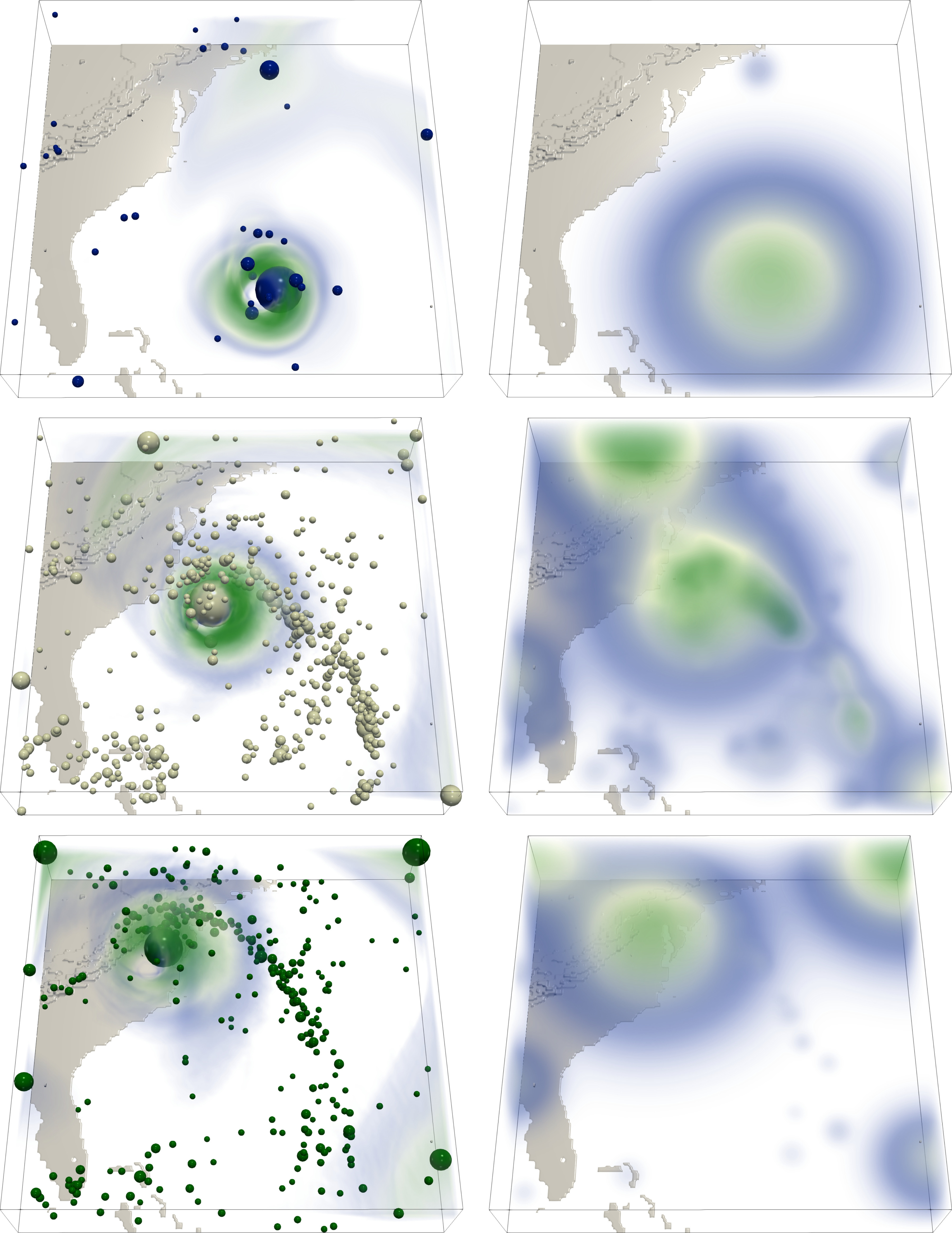}
 
 \includegraphics[width=\linewidth]{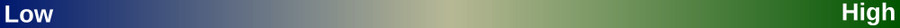}
 
 \vcaption{Three representative members for the \emph{Isabel} 
volumetric ensemble (left). Local maxima (scaled by 
persistence) of velocity magnitude capture the eyewall of the hurricane, 
high wind speed peripheral regions as well as the hurricane's tail (second and 
third 
row). The persistence maps (local maxima only, right) capture 
these 
subtle features by construction, while taking less into account noisy critical 
points (smallest spheres).
}
\label{fig_isabella_cp}
\end{figure}

\begin{figure}
  \centering
  \includegraphics[height=2.3cm]{colorMap_blueGreen.jpg}
  \hfill
  \includegraphics[height=2.3cm]{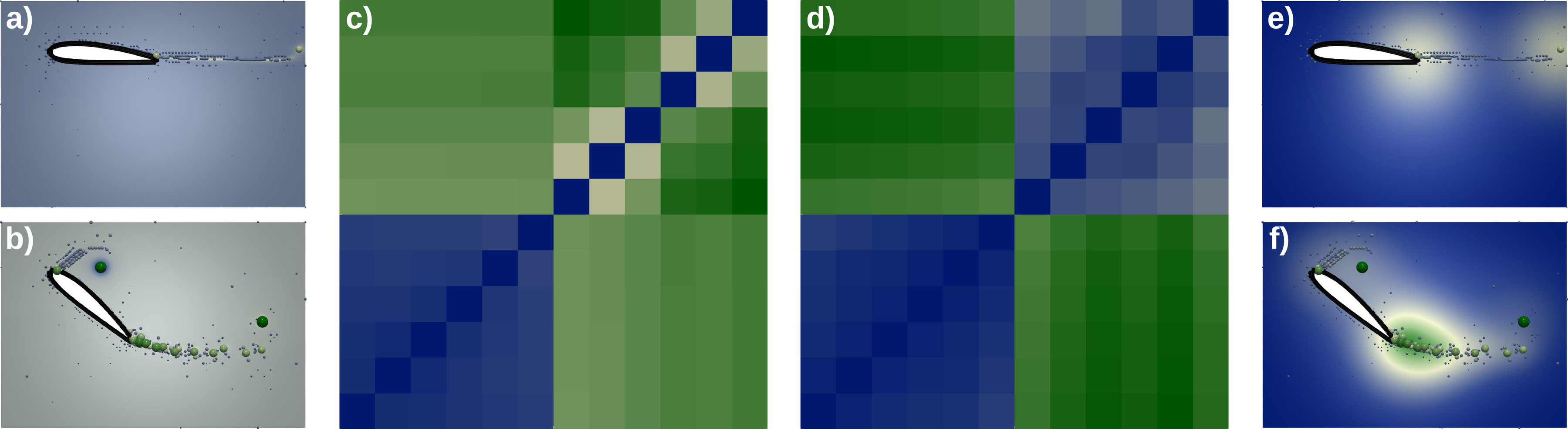}
 \vcaption{Two representative members of the \emph{Starting vortex} ensemble
((a) and (b)) along with their critical points scaled by persistence. Given the 
small spatial extent of these features, the $L_2$ norm between the actual data 
values fails at capturing similarities between members belonging to the 
bottom configuration (b), as denoted by the corresponding distance matrix (c), 
where distances are important (green) in the upper-right corner. In contrast, 
the 
distance matrix (d) computed from persistence maps ((e) and (f)) exhibits much 
smaller distances (blue) between these members, facilitating their gathering in 
the 
low dimensional space $\mathcal{P}$.}
 \label{fig_distanceMatrix}
\end{figure}

\begin{table}
  \centering
  \vcaption{Running time (in seconds, with 12 cores) for the different steps of 
  our approach: persistence maps ($P.M.$, \autoref{sec_persistenceMap_eq}), 
  distance matrix ($D.M.$, \autoref{sec_distanceMatrix}), 
  low-dimensional embedding ($E.$, \autoref{sec_eigenMap}),
  clustering ($C.$, \autoref{sec_subClustering}) and mandatory critical points 
($M.C.P.$, \autoref{sec_confidenceRegions}).}
  \label{tab_runningTimes}
  \scalebox{0.65}{
    \centering
    \begin{tabular}{l|rr|r|rrrr|r}
    \toprule
    Dataset &$n$ & $|\domain^0|$ &
    P.M. & 
D.M. & E. & C. & M.C.P. & \textbf{Total}\\
    \midrule
    Gaussians (\autoref{fig_overview}) & 100 & 262,144 
      & 57.28 
      & 1.03 
      & 0.67 
      & 0.08 
      & 2.53  
      & \textbf{61.59} \\
    Vortex street (\autoref{fig_teaser}) & 45 & 30,000 
      & 2.28 
      & 0.02 
      & 0.67 
      & 0.09 
      & 0.22  
      & \textbf{3.28} \\
    Starting vortex (\autoref{fig_startingVortex}) & 12 & 1,500,000 
      & 61.44 
      & 0.09
      & 0.65
      & 0.07
      & 9.08  
      & \textbf{71.33} \\
    Isabel (\autoref{fig_isabella}) & 12 & 3,125,000
      & 168.70
      & 0.18
      & 0.63 
      & 0.07
      & 41.84  
      & \textbf{211.68} \\
    Sea Surface Height (\autoref{fig_sea}) & 48 & 1,036,800
      & 290.25
      & 0.99
      & 0.65 
      & 0.08 
      & 8.38  
      & \textbf{300.35} \\
    \bottomrule
    \end{tabular}
  }
\end{table}

\begin{figure*}
  \vspace{-2ex}
\centering
  \includegraphics[height=2.75cm]{colorMap_blueGreen.jpg}
  \hspace{2ex}
 \includegraphics[height=2.75cm]{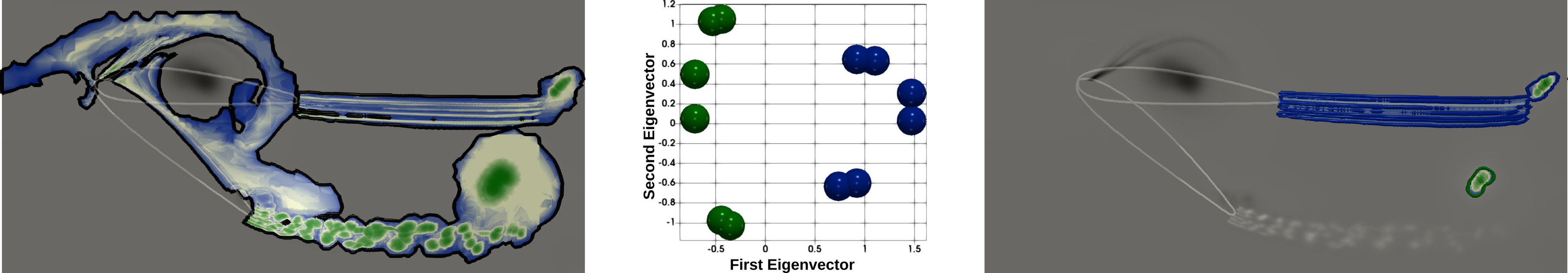}
 \vcaption{Mandatory critical point for the entire \emph{Starting vortex} 
ensemble \julien{}{(left)} and the corresponding 
\julien{likelihood of critical point appearance}{critical point positional 
variability estimation} (\julien{}{color map}, left). 
Given the trend variability of this ensemble, this global extraction identifies 
only one, very large, mandatory maximum (colored region) describing both 
regimes, although these two vortices never occur simultaneously in the data. 
The persistence atlas automatically identifies the two trends present in the 
data, as shown in the planar view (center), resulting in much more accurate 
predictions for the appearance of the distinctly identified vortices (green 
and blue regions, right).}
\label{fig_startingVortex}
\end{figure*}

\begin{figure*}
  \centering
  \includegraphics[height=3.95cm]{colorMap_blueGreen.jpg}
  \hspace{4ex}
 \includegraphics[height=3.95cm]{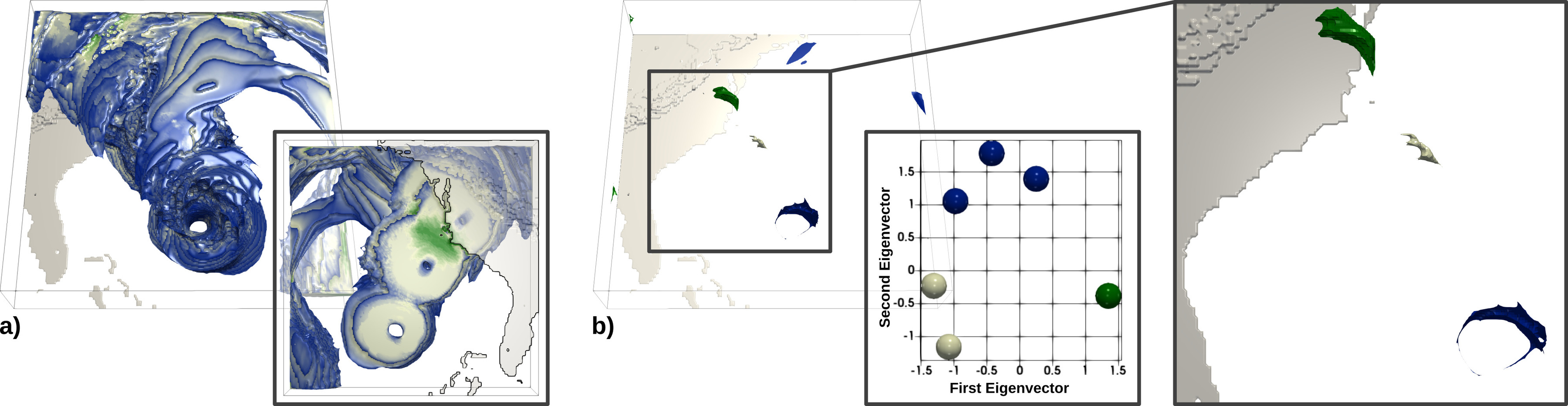}
 \vcaption{Mandatory critical point for the\julien{entire}{} \emph{Isabel} 
ensemble, 
viewed from above and below (a). Given the trend variability of this ensemble, 
this\julien{global}{}
extraction identifies only one, very large, mandatory maximum 
\julien{(colored region)}{(region colored by critical point positional 
variability)} merging the three 
distinct states of the hurricane 
(formation, drift and landfall). In contrast, the persistence atlas manages to 
isolate these three states and provides\julien{much}{} more accurate 
confidence regions 
for the position of the hurricane eyewall (colored regions).}
\label{fig_isabella}
\vspace{-10pt}
\end{figure*}

\subsection{Experiments}
Figures \ref{fig_teaser} and \ref{fig_vortexStreet_pm} to \ref{fig_sea} 
report various experiments on simulated and acquired 2D and 3D ensemble 
datasets. \autoref{fig_teaser} presents our entire approach on an 
ensemble of 45 von K\'arm\'an vortex streets, where the considered scalar 
data is the orthogonal component of the curl taken at a fixed time-step, for 
five different fluids of distinct viscosity (9 runs per fluid, each run with 
varying Reynolds numbers). 
\julien{}{For such scalar fields, local extrema
are typically considered as 
reliable estimations of the center of the vortices.}
\julien{The critical points}{Extrema}
of a few representative 
members 
(\autoref{fig_teaser}(a)) exhibit clearly distinct layout patterns, in terms of 
both the position and number of vortices, revealing high \emph{spatial} and 
\emph{trend} variabilities within the ensemble. The mandatory critical points 
estimated for the entire ensemble are particularly conservative given these 
variabilities: only one region is extracted for each side of the street (one 
for minima, one for maxima). The persistence atlas manages to automatically 
identify five clusters in the ensemble, corresponding to distinct critical 
point layouts (one per viscosity regime). The mandatory critical points 
extracted from these clusters provide more accurate and useful predictions for 
the appearance of vortices (colored regions in (d) to (h), one color per 
cluster). In particular, the persistence atlas reveals that the number of 
vortices increases with the Reynolds numbers (from left to right: 6, 10, 12, 14 
and 15 vortices) while the spatial variability of each vortex tends to decrease 
for increasing Reynolds numbers (smaller mandatory critical points).
\autoref{fig_vortexStreet_pm} illustrates persistence maps for five 
representative members of the ensemble and shows how salient features are 
captured by this representation.
\autoref{fig_isabella_cp} shows persistence 
maps on a volumetric ensemble composed of groups of key timesteps 
(formation, drift and landfall) in the simulation of the Isabel hurricane 
\cite{scivisIsabel}. 
\julien{}{For such datasets, the eyewall of the hurricane is typically 
characterized by high wind velocities (green regions, 
\autoref{fig_isabella_cp}, left) and contains salient maxima.}
In particular, this figure shows that subtle features of 
the hurricane (eyewall, high wind speed peripheral regions and hurricane's 
tail) are well captured by local maxima of the wind velocity magnitude and by 
the corresponding persistence maps. As discussed in 
\autoref{sec_distanceMatrix}, the $L_2$ norm between persistence maps is more 
suited to our purpose than the $L_2$ norm between the actual data values, since 
it is more robust to noise and global shifts in data values, while better 
discriminating changes in salient features (\autoref{fig_persistenceMap}).
\autoref{fig_distanceMatrix} further exemplifies this observation on the 
\emph{Starting vortex} ensemble, which includes 12 runs of a 2D
simulation of the formation of a vortex behind a wing, for two distinct wing 
configurations. The considered scalar field is the curl orthogonal 
component \julien{}{and salient extrema are expected at the center of 
vortices}. 
Given the 
small spatial extent of the features behind the wing, the $L_2$ norm between 
the actual data 
values fails at capturing similarities between members belonging to the 
second configuration, as denoted by the corresponding distance matrix, 
where distances are important in the upper-right corner. 
In particular, two members belonging to the same wing configuration are 
reported by this distance as the two furthest members (darkest green entry).
In contrast, the 
distance matrix computed from persistence maps exhibits much 
smaller (resp. higher) distances between the members belonging to a common 
(resp. distinct) wing configuration.

\autoref{fig_startingVortex} shows the persistence atlas for the \emph{Starting 
vortex} ensemble. Given the trend variability of this dataset, the mandatory 
critical points computed from the entire ensemble exhibit
only one, very large, mandatory maximum (colored region) describing the 
appareance of vortices for both wing configurations,
although these two vortices never occur simultaneously in the data. 
The persistence atlas automatically identifies the two trends present in the 
data, as shown in the planar view (center), resulting in much more accurate 
predictions for the appearance of the distinctly identified vortices (green 
and blue region, right).
\autoref{fig_isabella} shows the persistence atlas for 
the \emph{Isabel} ensemble. Similarly to the previous example, mandatory 
critical points computed from the entire ensemble identify only one, very 
large, mandatory maximum, which merges the three distinct states of the 
hurricane. In contrast, the persistence atlas manages to isolate these three 
states and provides much more accurate confidence regions for the position of 
the hurricane eyewall. Note that this example is the only dataset for which the 
initial automatic suggestion for the number of clusters $k$
provided by the eigengap heuristic needed 
adjustment. All the other results have been generated with the automatic 
suggestion.
\autoref{fig_sea} shows the persistence atlas for the \emph{Sea surface height} 
ensemble, which is composed of 48 observations taken in January, April, July 
and October 2012 
(\href{https://ecco.jpl.nasa.gov/products/all/}{https://ecco.jpl.nasa.gov/products/all/}).
\julien{}{For such datasets, salient extrema in the height are expected at the 
center of eddies.}
The mandatory critical points globally extracted on the entire ensemble identify
only few features (\autoref{fig_sea}(a)), due to the high pointwise data 
variability (\autoref{sec_mandatoryCriticalPoints}). The clustering 
automatically performed by our approach based on the persistence maps correctly 
identifies four clusters, corresponding to the four seasons: winter (c), spring 
(d), summer (e) and fall (f). This seasonal decomposition drastically 
reduces pointwise data variability and enables mandatory critical points to 
identify many more structures, corresponding to clockwise and counterwise 
vortices (minima and maxima) and revealing complex structures in the Gulf 
stream area (insets). Note that, for this example, due to the high number of 
critical points and their respective proximity, the parameter $\gamma$,
controlling the spread of salient features in the persistence map, has been set 
to $0.01$ instead of the default value ($0.1$).

\subsection{Time performance}
\label{sec_performance}
 \autoref{tab_runningTimes} presents the running times we obtained 
\brian{with}{for} the datasets presented in this paper.
The most time consuming \brian{part}{portion} of our approach is the computation of the 
persistence maps, which typically needs to be run for each ensemble member 
\brian{in}{as} a 
pre-process. Since the number of pairs in the diagram is typically proportional 
to the number $|\domain^0|$ of vertices in the domain, this part
requires $O(n\times|\domain^0|^2)$ steps overall.
%
In practice, to accelerate this computation, we 
ignore all pairs with a persistence less than $1\%$ of the total function 
range. 
The distance matrix computation 
takes $O(n^2 \times|\domain^0| )$ steps, but 
%
since $n$ is 
typically much smaller than $|\domain^0|$, the computation time for this step 
is small in practice. Both the spectral 
embedding (\autoref{sec_eigenMap}) and clustering (\autoref{sec_subClustering}) 
employ iterative solvers but these computations are typically the fastest steps 
of the pipeline. \julien{Finally, t}{T}he computation of the mandatory critical 
points for 
each cluster admits quadratic complexity  $O(|\domain^0|^2)$.

Most of these steps can be trivially parallelized.
The persistence diagram is computed in parallel \cite{gueunet_ldav17} and the 
persistence map can be evaluated independently for each vertex.
Each entry of the distance matrix $\Phi$  
(\autoref{sec_distanceMatrix}) can be computed independently.
Finally, mandatory critical points are computed in parallel for 
each cluster. As reported in \autoref{tab_runningTimes}, 
once the persistence maps have been computed in a pre-process, the rest 
of the framework is sufficiently fast to allow interactiv\julien{e 
exploration}{ity}.

\subsection{Comparison}
\begin{figure}
  \vspace{-2ex}
  \includegraphics[width=\linewidth]{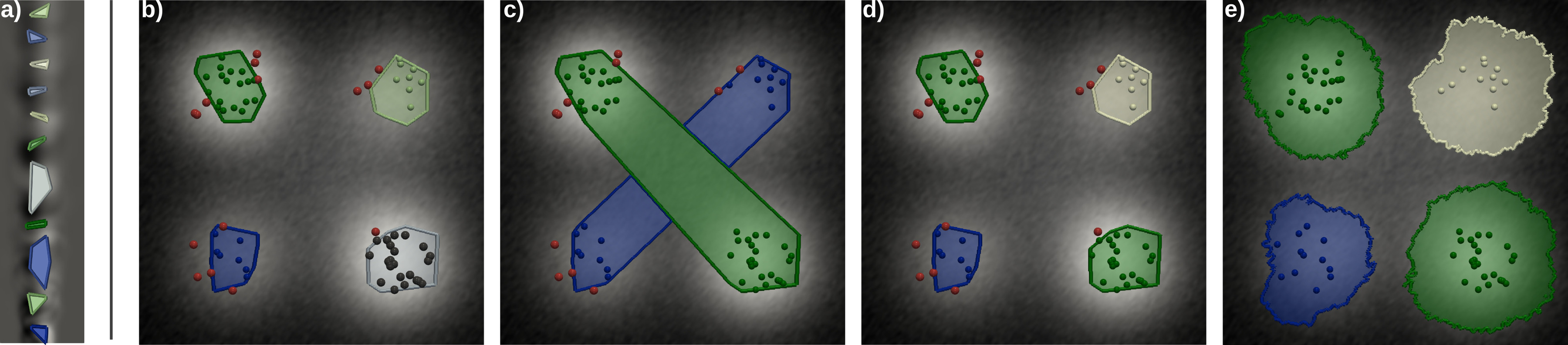}
  \vspace{-2ex}
  \vcaption{
  \julien{Comparison to a baseline strategy clustering critical points in the 
geometrical domain.}{Comparison to alternative critical point clustering 
strategies.}
\julien{(a) Features which never occur simultaneously in the ensemble but which 
are 
close from each other can be clustered together erroneously (dataset from 
\autoref{fig_teaser}).}{(a) Clustering the critical points directly in the 
spatial domain can cluster erroneously features which never occur simultaneously 
in the ensemble but which are close from each other in the spatial domain  
(dataset from 
\autoref{fig_teaser}).}
(b) On the contrary, features which always occur simultaneously 
(\autoref{fig_overview}) may not be clustered together if they are too distant 
from each other, leading to
an \julien{incorrect}{overestimated} number of clusters (black spheres). 
\julien{}{(c) Kernel based methods for persistence diagrams 
\cite{ReininghausHBK15, implemKernel} do not take spatial embedding into account
and cluster together members with the same persistence profile 
irrespective of the features' location, leading to an underestimated number 
of 
clusters.}
\julien{(c)}{(d)} 
Even when the clustering is explicitly \julien{provided}{given}, confidence 
regions based on 
convex hulls miss 21\% of the critical points of the test ensemble (red 
spheres). \julien{(d)}{(e)} The persistence atlas estimated from the training 
ensemble provides a correct prediction for 100\% 
of the critical points of the test ensemble (colored spheres).
\vspace{-1.5ex}}
  \label{fig_comparison}
\end{figure}

In this section, we compare our approach to 
\julien{
a \emph{baseline} approach, 
which consists in simply clustering persistent critical points in the 
geometrical domain, by using a vanilla implementation of spectral clustering, 
combined with our eigengap heuristic for the automatic suggestion 
of the number of clusters (\autoref{sec_subClustering}).}{alternative critical 
point clustering strategies. First, we consider a baseline approach, 
which consists in simply clustering persistent critical points in the 
spatial domain, by using a vanilla implementation of spectral clustering, 
combined with our eigengap heuristic for the automatic suggestion 
of the number of clusters (\autoref{sec_subClustering}).}
Once such clusters 
have been computed, this baseline strategy evaluates confidence regions for the 
appearance of critical points by considering the convex hull of each cluster 
in the \julien{geometrical}{spatial} domain.
As shown in \autoref{fig_comparison}(a), this simple strategy provides 
unsatisfactory results for the von K\'arm\'an vortex street ensemble 
(\autoref{fig_teaser}) since features which never occur simultaneously in the 
ensemble are clustered based on their proximity. In particular, the extracted 
clusters mix the two types of vortices (right and left) and group them based 
on their distance from the obstacle (bottom).

\julien{To compare this 
strategy to the persistence atlas,}{To further evaluate our approach,} we 
consider the ensemble from 
\autoref{fig_overview}, which we split in half\brian{,}{} into a training and \brian{a }{}test 
ensemble. The training ensemble is analyzed with \julien{}{\emph{(i)}} the 
baseline \julien{strategy}{approach}
(\autoref{fig_comparison}(b))\julien{}{, \julien{}{\emph{(ii)}} a strategy 
based on the kernel method 
by Reininghaus et al. \cite{ReininghausHBK15} 
(\autoref{fig_comparison}(c), where the distance matrix considered for 
clustering has been generated 
with the authors' implementation of the kernel method 
\cite{implemKernel} run with default parameters)} and 
\julien{}{\emph{(iii)}}
the persistence atlas 
(\autoref{fig_comparison}(\julien{d}{e})). To quantitatively evaluate the 
prediction 
performance of \julien{the two}{these} approaches, we consider the persistent 
critical points of 
the test ensemble having a persistence higher than 20\% of the function range.
Next, the test critical points are 
assigned to the confidence region in which they land in the domain 
(spheres of matching colors in \autoref{fig_comparison}). As shown in  
\autoref{fig_comparison}(b), 
the baseline \julien{strategy}{approach}
\julien{yields an incorrect}{overestimates the} number of clusters. 
\julien{Moreover}{In particular}, 
it 
fails at clustering together features which always occur 
simultaneously
(dark green and light blue clusters in \autoref{fig_comparison}(b)). 
\julien{}{Kernel based methods for persistence diagrams \cite{ReininghausHBK15, 
implemKernel} do not take the spatial embedding of 
critical points into account (\autoref{fig_comparison}(c)) and 
cluster members with the same persistence profile, irrespective of 
the features' location. This leads to an 
underestimated number of clusters: the blue and white clusters of 
\autoref{fig_overview}, which both include a single very persistent maximum,
are erroneously merged although the corresponding features never occur 
simultaneously in the ensemble.
Moreover, convex hulls obtained from this clustering overestimate the size of 
the confidence regions in the presence of multiple salient features per 
cluster.}
Even
when the correct clustering is explicitly provided 
(\autoref{fig_comparison}\julien{(c)}{(d)}),
confidence regions based on convex hulls 
miss 
21\% of the persistent critical points of the test ensemble. In contrast, the 
persistence atlas \julien{}{(\autoref{fig_comparison}(e))} 
provides a correct prediction for 100\% of the critical points 
of the test ensemble, which  illustrates the quantitative 
performance of the persistence atlas regarding critical point prediction.

\begin{figure}
  \vspace{-2ex}
  \centering
 \includegraphics[width=0.9\linewidth]{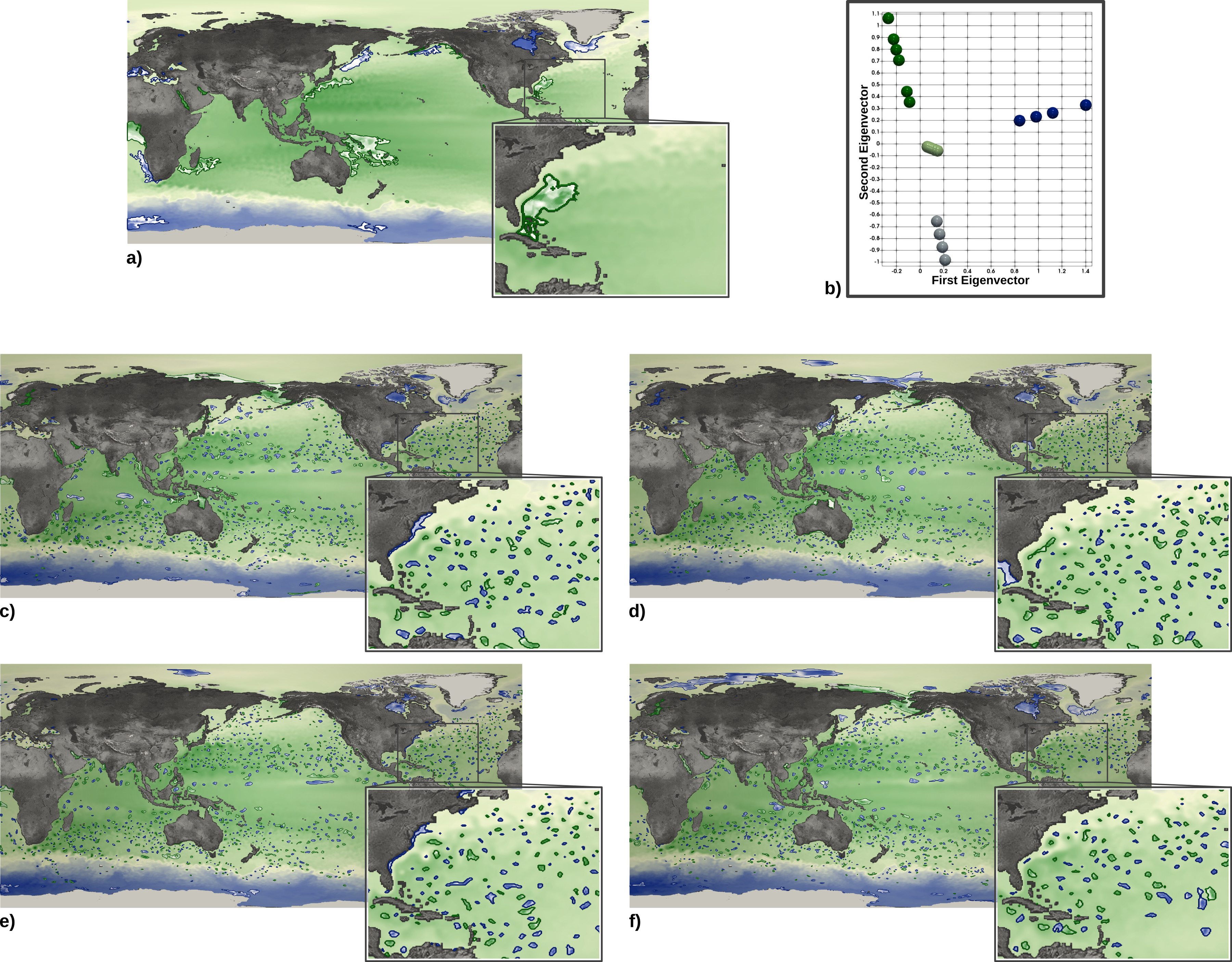}
  
 \includegraphics[width=\linewidth]{colorMap_blueGreen_h.jpg}
 \vspace{-3ex}
 \vcaption{Persistence atlas for the \emph{Sea surface height} ensemble.
 (a)  Mandatory critical points for the entire ensemble miss many features and 
over-estimate the spatial variability of the extracted structures. (b) 
The planar layout of the space of persistence maps shows the automatically 
evaluated clustering of the ensemble (one color per cluster), which correspond 
to the four seasons.
(c), (d), (e) and (f) The mandatory critical points for each of the 
 identified clusters (respectively: winter, spring, summer and 
fall) provide a more precise critical point variability estimation, revealing 
hundreds of vortices (blue: minima, green: maxima).\vspace{-1.75ex}}
 \label{fig_sea}
\end{figure}
\subsection{Limitations}
\label{sec_limitations}
\julien{}{Our entire pipeline assumes that the input data is given as a 
collection of piecewise linear scalar fields (\autoref{sec:background}). In 
many applications~\cite{ScheuermannTH99}, this may be too restrictive 
(motivating taylored interpolants for uncertainty modeling 
\cite{SchlegelKS12}). However, generalizing the TDA arsenal to a larger set of 
interpolants is a vast research topic (see \cite{NuchaBHN17} for an example) 
which goes beyond the scope of this paper.}
Our approach focuses on $(0, 1)$ and $\big((d-1), d\big))$ persistence pairs, 
which correspond to pairs only involving minima and maxima\brian{, and 
p}{. P}ersistence 
maps (\equref{eq_persistenceMap}) are \brian{}{therefore} only computed based 
on 
the location of \brian{}{either} the minima or maxima 
(Figs.~\ref{fig_overview}, 
\ref{fig_startingVortex}
and \ref{fig_isabella_cp})\brian{}{,} or 
both (Figs.~\ref{fig_vortexStreet_pm} and \ref{fig_sea}). 
\BrianSays{doublecheck the change} Hence, saddle points 
are not taken into account by our framework in its current form. However, 
we \brian{}{have}
found that in practice the correspondence between saddle points and features of 
interest was less clear in \brian{}{our} applications. Also, when the data 
exhibits salient large
flat plateaus, persistent critical points can appear in arbitrary locations 
inside these plateaus\brian{, }{. This can} potentially impair\brian{ing}{} the 
stability of the persistence 
map. However, we did not observe this behavior in practice on our datasets as 
large plateaus, when they occurred, were not collocated with salient features.
We found in practice that using constant weights ($W_{xy} = 1$) for the 
evaluation of the Laplacian of the adjacency graph of the persistence maps
(\autoref{sec_clustering}) resulted in more stable and accurate clusterings 
than the second weighting strategy (based on heat kernels) described by Belkin 
et al. \cite{belkin2003laplacian}. However, constant weights result in the 
limitation that several members can be projected to the exact same point in the 
low dimensional space when they exhibit a very similar neighborhood pattern in 
the adjacency graph. \brian{Then}{In this case,} the number of visible points in
our planar layouts 
may be smaller than the actual number of members. However, this non-uniqueness 
in the embedding only occurs for persistence maps which are very close \brian{from}{to} 
each other, hence it does not impact negatively the clustering \brian{}{or 
analysis}.
\julien{Finally, a}{A}lthough the automatic suggestion for the number of 
clusters $k$ 
provided satisfactory results for all but one example (where it needed to be 
changed from $2$ to $3$, \autoref{fig_isabella}), an exhaustive interactive 
exploration may be needed when there is no clear trend in the ensemble.
\julien{}{Finally, the persistence atlas currently displays simultaneously 
mandatory critical points for all clusters. This may result in cluttered 
visualizations due to overlapping. Although we provide users with the 
possibility of refining this visualization to a selected subset of clusters, 
improved strategies for the overall visualization of the atlas could be 
researched in the future.}

\section{Conclusion}
In this paper, we presented the \emph{Persistence Atlas}, an approach for the 
visual analysis of the spatial variability of features of interest represented 
by critical points in ensemble data. By analyzing the structure of the ensemble 
in terms of patterns of critical point layouts, our method addresses 
\emph{trend} variability, by identifying clusters of ensemble members which 
share a common geometrical configuration of critical points. By computing 
mandatory critical points for each cluster, our approach addresses 
\emph{spatial} variability, by showing minimal regions where at least 
one critical point is guaranteed to occur for each member of the cluster, hence 
conveying the 
 possible 
extent of features 
for each trend.
Our approach is based on the new notion of \emph{Persistence Map}, which 
describes the local density in critical points and leverages topological 
persistence to emphasize salient features, 
and which has been shown to be well suited for the
purpose of comparing 
geometrical layouts of critical points. We showed how to leverage spectral 
embedding methods to provide low-dimensional views representing the main trends 
found in the ensemble. We also showed how to leverage spectral clustering to 
automatically identify revelant clusters of ensemble members and how to provide 
relevant automatic guesses based on eigengaps for the number of clusters.
In practice, our approach has been shown to provide more accurate descriptions 
of the variability of critical points than global methods, such as the original 
mandatory critical points \cite{tierny_ev14}, which 
either miss features or
considerably over-estimate spatial variability 
in the presence of trend variability. 
\julien{Moreover, in contrast to these methods, 
our technique manages to reveal, 
thanks to its clustering capabilities,
subtle dependences in the occurrence of features of interest, by separating in 
different clusters features which never occur simultaneously in the ensemble 
members.}{} We quantitatively evaluated the prediction accuracy of our method 
and 
showed that it compared favorably to a baseline strategy based on an
off-the-shelf clustering approach. 
\julien{From a performance point of view, 
the computation of persistence maps can be time consuming. However, this step 
can be run as a pre-process and the rest of our framework can be run at 
interactive rates, allowing interactive exploration.}{}
Our work extends recent advances in the visual analysis of spatial variability 
in ensembles of geometrical objects, such as level sets \cite{FerstlKRW16} 
or streamlines \cite{FerstlBW16}, to topological structures. In particular, we 
focused in this paper on features of interest represented by critical points.
However, many more topological constructions 
could benefit from a similar variability analysis based on such tailored 
clustering strategies. For instance, the separatrices of the Morse-Smale 
complex \cite{gyulassy_vis08, robins11} have been shown to excel at 
representing filament structures in various applications, such as chemistry 
\cite{chemistry_vis14, harshChemistry} or astrophysics \cite{sousbie11, 
shivashankar2016felix}, and studying their trend and spatial variabilities 
would be of tremedous help for the understanding of non-deterministic models in 
these applications. 
By first focusing on critical points, we believe we made a 
first step in this direction, which will be helpful and inspirational for 
future generalizations to other topological constructions.

%
%
%
%
%
%
%
%
%

\acknowledgments{
\vspace{-.5ex}
\footnotesize{
This work is partially supported by the BPI grant ``AVIDO'' (PIA
FSN2, reference P112017-2661376/DOS0021427), 
NSF CRII 1657020, and NSF/NIH QuBBD 1664848.
We would like to 
thank the reviewers for their thoughtful remarks and 
suggestions. Julien Tierny 
would like to 
dedicate this paper to his son Otis.}
}

\clearpage

\bibliographystyle{abbrv-doi}

\bibliography{paper}

\begin{thebibliography}{10}

\bibitem{GUM}
{ISO/IEC} {G}uide 98-3:2008 uncertainty of measurement-part 3: Guide to the
  expression of uncertainty in measurement ({GUM}), 2008.

\bibitem{surveyPCA}
H.~Abdi and L.~Williams.
\newblock Principal component analysis.
\newblock {\em Wiley Interdisciplinary Reviews: Computational Statistics},
  2010.

\bibitem{persistenceImages}
H.~Adams, T.~Emerson, M.~Kirby, R.~Neville, C.~Peterson, P.~Shipman,
  S.~Chepushtanova, E.~Hanson, F.~Motta, and L.~Ziegelmeier.
\newblock Persistence images: A stable vector representation of persistent
  homology.
\newblock {\em Journal of Machine Learning Research}, 2017.

\bibitem{AthawaleE13}
T.~Athawale and A.~Entezari.
\newblock Uncertainty quantification in linear interpolation for isosurface
  extraction.
\newblock {\em IEEE Transactions on Visualization and Computer Graphics (Proc.
  of IEEE VIS)}, 2013.

\bibitem{AthawaleSE16}
T.~Athawale, E.~Sakhaee, and A.~Entezari.
\newblock Isosurface visualization of data with nonparametric models for
  uncertainty.
\newblock {\em IEEE Transactions on Visualization and Computer Graphics (Proc.
  of IEEE VIS)}, 2016.

\bibitem{banchoff70}
T.~F. Banchoff.
\newblock Critical points and curvature for embedded polyhedral surfaces.
\newblock {\em The American Mathematical Monthly}, 1970.

\bibitem{bauer14}
U.~Bauer, X.~Ge, and Y.~Wang.
\newblock Measuring distance between {R}eeb graphs.
\newblock In {\em Symp. on Comp. Geom.}, 2014.

\bibitem{BeketayevYMWH14}
K.~Beketayev, D.~Yeliussizov, D.~Morozov, G.~H. Weber, and B.~Hamann.
\newblock Measuring the distance between merge trees.
\newblock In {\em Topological Methods in Data Analysis and Visualization III,
  Theory, Algorithms, and Applications}. 2014.

\bibitem{belkin2003laplacian}
M.~Belkin and P.~Niyogi.
\newblock Laplacian eigenmaps for dimensionality reduction and data
  representation.
\newblock {\em Neural Computation}, 2003.

\bibitem{harshChemistry}
H.~Bhatia, A.~G. Gyulassy, V.~Lordi, J.~E. Pask, V.~Pascucci, and P.-T. Bremer.
\newblock Topoms: Comprehensive topological exploration for molecular and
  condensed-matter systems.
\newblock {\em Journal of Computational Chemistry}, 2018.

\bibitem{bhatia12}
H.~Bhatia, S.~Jadhav, P.~Bremer, G.~Chen, J.~Levine, L.~Nonato, and
  V.~Pascucci.
\newblock Flow visualization with quantified spatial and temporal errors using
  edge maps.
\newblock {\em IEEE Transactions on Visualization and Computer Graphics},
  18(9):1383--1396, 2012.

\bibitem{bonneau:hal-01060465}
G.-P. Bonneau, H.-C. Hege, C.~Johnson, M.~M. Oliveira, K.~Potter, and
  P.~Rheingans.
\newblock {Overview and State-of-the-Art of Uncertainty Visualization}.
\newblock In {\em {Scientific Visualization: Uncertainty, Multifield,
  Biomedical, Scalable}}, Mathematics and Visualization. {Springer}, 2014.

\bibitem{borg05}
I.~Borg and P.~Groenen.
\newblock {\em Modern multidimensional scaling: Theory and Applications}.
\newblock Springer, 2005.

\bibitem{botnan18}
M.~B. Botnan and H.~B. Bjerkevik.
\newblock Computational complexity of the interleaving distance.
\newblock In {\em Symp. on Comp. Geom.}, 2018.

\bibitem{bremer_tvcg11}
P.~Bremer, G.~Weber, J.~Tierny, V.~Pascucci, M.~Day, and J.~Bell.
\newblock Interactive exploration and analysis of large scale simulations using
  topology-based data segmentation.
\newblock {\em IEEE Transactions on Visualization and Computer Graphics}, 2011.

\bibitem{DBLP:journals/cgf/BrucknerM10}
S.~Bruckner and T.~M{\"{o}}ller.
\newblock Isosurface similarity maps.
\newblock {\em Computer Graphics Forum (Proc. of EuroVis)}, 2010.

\bibitem{CarriereCO17}
M.~Carri{\`{e}}re, M.~Cuturi, and S.~Oudot.
\newblock Sliced wasserstein kernel for persistence diagrams.
\newblock In {\em ICML}, 2017.

\bibitem{interleaving_distance}
F.~Chazal, D.~Cohen{-}Steiner, M.~Glisse, L.~J. Guibas, and S.~Oudot.
\newblock Proximity of persistence modules and their diagrams.
\newblock In {\em Symp. on Comp. Geom.}, 2009.

\bibitem{chazal13}
F.~Chazal, L.~Guibas, S.~Oudot, and P.~Skraba.
\newblock Persistence-based clustering in {R}iemannian manifolds.
\newblock {\em Journal of the ACM}, 2013.

\bibitem{cohen-steiner05}
D.~Cohen-Steiner, H.~Edelsbrunner, and J.~Harer.
\newblock Stability of persistence diagrams.
\newblock In {\em Symp. on Comp. Geom.}, 2005.

\bibitem{diggle02}
P.~Diggle, P.~Heagerty, K.-Y. Liang, and S.~Zeger.
\newblock {\em Analysis of longitudinal data}.
\newblock Oxford University Press, 2002.

\bibitem{edelsbrunner09}
H.~Edelsbrunner and J.~Harer.
\newblock {\em Computational Topology: An Introduction}.
\newblock American Mathematical Society, 2009.

\bibitem{edelsbrunner02}
H.~Edelsbrunner, D.~Letscher, and A.~Zomorodian.
\newblock Topological persistence and simplification.
\newblock {\em Disc. Compu. Geom.}, 2002.

\bibitem{edelsbrunner90}
H.~Edelsbrunner and E.~P. Mucke.
\newblock Simulation of simplicity: a technique to cope with degenerate cases
  in geometric algorithms.
\newblock {\em ACM Trans. on Graph.}, 1990.

\bibitem{favelier16}
G.~Favelier, C.~Gueunet, and J.~Tierny.
\newblock Visualizing ensembles of viscous fingers.
\newblock In {\em IEEE SciVis Contest}, 2016.

\bibitem{FerstlBW16}
F.~Ferstl, K.~B{\"{u}}rger, and R.~Westermann.
\newblock Streamline variability plots for characterizing the uncertainty in
  vector field ensembles.
\newblock {\em IEEE Transactions on Visualization and Computer Graphics (Proc.
  of IEEE VIS)}, 2016.

\bibitem{FerstlKRW16}
F.~Ferstl, M.~Kanzler, M.~Rautenhaus, and R.~Westermann.
\newblock Visual analysis of spatial variability and global correlations in
  ensembles of iso-contours.
\newblock {\em Computer Graphics Forum (Proc. of EuroVis)}, 2016.

\bibitem{chemistry_vis14}
D.~Guenther, R.~Alvarez-Boto, J.~Contreras-Garcia, J.-P. Piquemal, and
  J.~Tierny.
\newblock Characterizing molecular interactions in chemical systems.
\newblock {\em IEEE Transactions on Visualization and Computer Graphics (Proc.
  of IEEE VIS)}, 2014.

\bibitem{tierny_ev14}
D.~Guenther, J.~Salmon, and J.~Tierny.
\newblock Mandatory critical points of 2{D} uncertain scalar fields.
\newblock {\em Computer Graphics Forum (Proc. of EuroVis)}, 2014.

\bibitem{gueunet_ldav17}
C.~Gueunet, P.~Fortin, J.~Jomier, and J.~Tierny.
\newblock Task-based {A}ugmented {M}erge {T}rees with {F}ibonacci {H}eaps,.
\newblock In {\em IEEE LDAV}, 2017.

\bibitem{gyulassy_ev14}
A.~Gyulassy, P.~Bremer, R.~Grout, H.~Kolla, J.~Chen, and V.~Pascucci.
\newblock Stability of dissipation elements: A case study in combustion.
\newblock {\em Computer Graphics Forum (Proc. of EuroVis)}, 2014.

\bibitem{gyulassy_vis08}
A.~Gyulassy, P.~T. Bremer, B.~Hamann, and V.~Pascucci.
\newblock A practical approach to morse-smale complex computation: Scalability
  and generality.
\newblock {\em IEEE Transactions on Visualization and Computer Graphics (Proc.
  of IEEE VIS)}, 2008.

\bibitem{gyulassy_vis07}
A.~Gyulassy, M.~A. Duchaineau, V.~Natarajan, V.~Pascucci, E.~Bringa,
  A.~Higginbotham, and B.~Hamann.
\newblock Topologically clean distance fields.
\newblock {\em IEEE Transactions on Visualization and Computer Graphics (Proc.
  of IEEE VIS)}, 2007.

\bibitem{gyulassy_vis15}
A.~Gyulassy, A.~Knoll, K.~Lau, B.~Wang, P.~Bremer, M.~Papka, L.~A. Curtiss, and
  V.~Pascucci.
\newblock Interstitial and interlayer ion diffusion geometry extraction in
  graphitic nanosphere battery materials.
\newblock {\em IEEE Transactions on Visualization and Computer Graphics (Proc.
  of IEEE VIS)}, 2015.

\bibitem{heine16}
C.~Heine, H.~Leitte, M.~Hlawitschka, F.~Iuricich, L.~De~Floriani,
  G.~Scheuermann, H.~Hagen, and C.~Garth.
\newblock A survey of topology-based methods in visualization.
\newblock {\em Comp. Grap. For.}, 2016.

\bibitem{hilaga:sig:2001}
M.~Hilaga, Y.~Shinagawa, T.~Kohmura, and T.~L. Kunii.
\newblock Topology matching for fully automatic similarity estimation of 3{D}
  shapes.
\newblock In {\em Proc. of ACM SIGGRAPH}, 2001.

\bibitem{DBLP:journals/tvcg/HummelOGJ13}
M.~Hummel, H.~Obermaier, C.~Garth, and K.~I. Joy.
\newblock Comparative visual analysis of lagrangian transport in {CFD}
  ensembles.
\newblock {\em IEEE Transactions on Visualization and Computer Graphics (Proc.
  of IEEE VIS)}, 2013.

\bibitem{johnson03}
C.~Johnson and A.~Sanderson.
\newblock A next step: Visualizing errors and uncertainty.
\newblock {\em IEEE Computer Graphics and Applications}, 2003.

\bibitem{kasten_tvcg11}
J.~Kasten, J.~Reininghaus, I.~Hotz, and H.~Hege.
\newblock Two-dimensional time-dependent vortex regions based on the
  acceleration magnitude.
\newblock {\em IEEE Transactions on Visualization and Computer Graphics}, 2011.

\bibitem{implemKernel}
R.~Kwitt.
\newblock Persistence learning.
\newblock \url{https://github.com/rkwitt/persistence-learning}, 2015.

\bibitem{laney_vis06}
D.~E. Laney, P.~Bremer, A.~Mascarenhas, P.~Miller, and V.~Pascucci.
\newblock Understanding the structure of the turbulent mixing layer in
  hydrodynamic instabilities.
\newblock {\em IEEE Transactions on Visualization and Computer Graphics (Proc.
  of IEEE VIS)}, 2006.

\bibitem{LiebmannS16}
T.~Liebmann and G.~Scheuermann.
\newblock Critical points of gaussian-distributed scalar fields on simplicial
  grids.
\newblock {\em Computer Graphics Forum (Proc. of EuroVis)}, 2016.

\bibitem{lloyd57}
S.~P. Lloyd.
\newblock Least square quantization in pcm.
\newblock Technical report, Bell Telephone Laboratories, 1957.

\bibitem{MacEachren05}
A.~M. MacEachren, A.~Robinson, S.~Hopper, S.~Gardner, R.~Murray, M.~Gahegan,
  and E.~Hetzler.
\newblock Visualizing geospatial information uncertainty: What we know and what
  we need to know.
\newblock {\em Cartography and Geographic Information Science}, 32(3):139--160,
  2005.

\bibitem{historicKmeans}
J.~B. MacQueen.
\newblock Some methods for classification and analysis of multivariate
  observations.
\newblock In {\em Proc. Symposium on Mathematical Statistics and Probability},
  1967.

\bibitem{milnor63}
J.~Milnor.
\newblock {\em Morse Theory}.
\newblock Princeton U. Press, 1963.

\bibitem{DBLP:journals/tvcg/MirzargarWK14}
M.~Mirzargar, R.~T. Whitaker, and R.~M. Kirby.
\newblock Curve boxplot: Generalization of boxplot for ensembles of curves.
\newblock {\em IEEE Transactions on Visualization and Computer Graphics (Proc.
  of IEEE VIS)}, 2014.

\bibitem{munkres}
J.~Munkres.
\newblock Algorithms for the assignment and transportation problems.
\newblock {\em Journal of the Society for Industrial and Applied Mathematics},
  1957.

\bibitem{NuchaBHN17}
G.~Nucha, G.~Bonneau, S.~Hahmann, and V.~Natarajan.
\newblock Computing contour trees for 2d piecewise polynomial functions.
\newblock {\em Computer Graphics Forum (Proc. of EuroVis)}, 2017.

\bibitem{DBLP:journals/tvcg/OeltzeLKJTP14}
S.~Oeltze, D.~J. Lehmann, A.~Kuhn, G.~Janiga, H.~Theisel, and B.~Preim.
\newblock Blood flow clustering and applications invirtual stenting of
  intracranial aneurysms.
\newblock {\em IEEE Transactions on Visualization and Computer Graphics (Proc.
  of IEEE VIS)}, 2014.

\bibitem{otto10}
M.~Otto, T.~Germer, H.-C. Hege, and H.~Theisel.
\newblock Uncertain 2{D} vector field topology.
\newblock {\em Comp. Graph. For.}, 29:347--356, 2010.

\bibitem{OttoGT11}
M.~Otto, T.~Germer, and H.~Theisel.
\newblock Uncertain topology of 3d vector fields.
\newblock In {\em Proc. of IEEE PacificVis}, 2011.

\bibitem{Pang97}
A.~T. Pang, C.~M. Wittenbrink, and S.~K. Lodha.
\newblock Approaches to uncertainty visualization.
\newblock {\em The Visual Computer}, 13(8):370--390, 1997.

\bibitem{scikitlearn}
F.~Pedregosa, G.~Varoquaux, A.~Gramfort, V.~Michel, B.~Thirion, O.~Grisel,
  M.~Blondel, P.~Prettenhofer, R.~Weiss, V.~Dubourg, J.~VanderPlas, A.~Passos,
  D.~Cournapeau, M.~Brucher, M.~Perrot, and E.~Duchesnay.
\newblock Scikit-learn: Machine learning in python.
\newblock {\em Journal of Machine Learning Research}, 2011.

\bibitem{pelleg00}
D.~Pelleg and A.~W. Moore.
\newblock X-means: Exteding k-means with efficient estimation of the number of
  clusters.
\newblock In {\em Proc. of ICML}, 2000.

\bibitem{petz12}
C.~Petz, K.~P\"othkow, and H.-C. Hege.
\newblock Probabilistic local features in uncertain vector fields with spatial
  correlation.
\newblock {\em Computer Graphics Forum (Proc. of EuroVis)},
  31(3pt2):1045--1054, 2012.

\bibitem{DBLP:journals/tvcg/PfaffelmoserMW13}
T.~Pfaffelmoser, M.~Mihai, and R.~Westermann.
\newblock Visualizing the variability of gradients in uncertain 2d scalar
  fields.
\newblock {\em IEEE Transactions on Visualization and Computer Graphics}, 2013.

\bibitem{pfaffelmoser11}
T.~Pfaffelmoser, M.~Reitinger, and R.~Westermann.
\newblock Visualizing the positional and geometrical variability of isosurfaces
  in uncertain scalar fields.
\newblock {\em Computer Graphics Forum (Proc. of EuroVis)}, 30:951--960, 2011.

\bibitem{DBLP:journals/cgf/PfaffelmoserW12}
T.~Pfaffelmoser and R.~Westermann.
\newblock Visualization of global correlation structures in uncertain 2d scalar
  fields.
\newblock {\em Computer Graphics Forum (Proc. of EuroVis)}, 2012.

\bibitem{pfaffelmoser13}
T.~Pfaffelmoser and R.~Westermann.
\newblock Visualizing contour distributions in 2d ensemble data.
\newblock In {\em EuroVis-Short Papers}, pp. 55--59. The Eurographics
  Association, 2013.

\bibitem{PhillipsWZ15}
J.~M. Phillips, B.~Wang, and Y.~Zheng.
\newblock Geometric inference on kernel density estimates.
\newblock In {\em Symp. on Comp. Geom.}, 2015.

\bibitem{poetkow11}
K.~P\"othkow and H.-C. Hege.
\newblock Positional uncertainty of isocontours: Condition analysis and
  probabilistic measures.
\newblock {\em IEEE Transactions on Visualization and Computer Graphics},
  17(10):1393--1406, 2011.

\bibitem{poetkow13}
K.~P\"othkow and H.-C. Hege.
\newblock Nonparametric models for uncertainty visualization.
\newblock {\em Computer Graphics Forum (Proc. of EuroVis)}, 32:131--140, 2013.

\bibitem{poethkow13b}
K.~P\"othkow, C.~Petz, and H.-C. Hege.
\newblock Approximate level-crossing probabilities for interactive
  visualization of uncertain isocontours.
\newblock {\em Int. J. Uncert. Quantif.}, 3:101--117, 2013.

\bibitem{DBLP:journals/cgf/PothkowWH11}
K.~P{\"{o}}thkow, B.~Weber, and H.~Hege.
\newblock Probabilistic marching cubes.
\newblock {\em Computer Graphics Forum (Proc. of EuroVis)}, 2011.

\bibitem{potter13}
K.~Potter, S.~Gerber, and E.~Anderson.
\newblock Visualization of uncertainty without a mean.
\newblock {\em IEEE CGA}, 33:75--79, 2013.

\bibitem{potter12}
K.~Potter, P.~Rosen, and C.~R. Johnson.
\newblock From quantification to visualization: A taxonomy of uncertainty
  visualization approaches.
\newblock In {\em Uncertainty Quantification in Scientific Computing}, vol.
  377, pp. 226--249. Springer, 2012.

\bibitem{DBLP:conf/icdm/PotterWBWDPJ09}
K.~Potter, A.~T. Wilson, P.~Bremer, D.~N. Williams, C.~M. Doutriaux,
  V.~Pascucci, and C.~R. Johnson.
\newblock Ensemble-vis: {A} framework for the statistical visualization of
  ensemble data.
\newblock In {\em {IEEE} International Conference on Data Mining Workshops},
  2009.

\bibitem{DBLP:journals/tvcg/QuinanM16}
P.~S. Quinan and M.~D. Meyer.
\newblock Visually comparing weather features in forecasts.
\newblock {\em IEEE Transactions on Visualization and Computer Graphics}, 2016.

\bibitem{ReininghausHBK15}
J.~Reininghaus, S.~Huber, U.~Bauer, and R.~Kwitt.
\newblock A stable multi-scale kernel for topological machine learning.
\newblock In {\em {IEEE} {CVPR}}, 2015.

\bibitem{robins11}
V.~Robins, P.~Wood, and A.~Sheppard.
\newblock Theory and algorithms for constructing discrete morse complexes from
  grayscale digital images.
\newblock {\em IEEE Trans. on Pat. Ana. and Mach. Int.}, 2011.

\bibitem{SaikiaSW14_branch_decomposition_comparison}
H.~Saikia, H.~Seidel, and T.~Weinkauf.
\newblock Extended branch decomposition graphs: Structural comparison of scalar
  data.
\newblock {\em Computer Graphics Forum (Proc. of EuroVis)}, 2014.

\bibitem{DBLP:journals/tvcg/SanyalZDMAM10}
J.~Sanyal, S.~Zhang, J.~Dyer, A.~Mercer, P.~Amburn, and R.~J. Moorhead.
\newblock Noodles: {A} tool for visualization of numerical weather model
  ensemble uncertainty.
\newblock {\em IEEE Transactions on Visualization and Computer Graphics (Proc.
  of IEEE VIS)}, 2010.

\bibitem{ScheuermannTH99}
G.~Scheuermann, X.~Tricoche, and H.~Hagen.
\newblock C1-interpolation for vector field topology visualization.
\newblock In {\em {IEEE} VIS}, 1999.

\bibitem{SchlegelKS12}
S.~Schlegel, N.~Korn, and G.~Scheuermann.
\newblock On the interpolation of data with normally distributed uncertainty
  for visualization.
\newblock {\em IEEE Transactions on Visualization and Computer Graphics (Proc.
  of IEEE VIS)}, 2012.

\bibitem{scivisIsabel}
I.~SciVisContest.
\newblock Simulation of the isabel hurricane.
\newblock \url{http://sciviscontest-staging.ieeevis.org/2004/data.html}.

\bibitem{DBLP:conf/cvpr/ShiM97}
J.~Shi and J.~Malik.
\newblock Normalized cuts and image segmentation.
\newblock In {\em Proc. of IEEE CVPR}, 1997.

\bibitem{shivashankar2016felix}
N.~Shivashankar, P.~Pranav, V.~Natarajan, R.~van~de Weygaert, E.~P. Bos, and
  S.~Rieder.
\newblock Felix: A topology based framework for visual exploration of cosmic
  filaments.
\newblock {\em IEEE Transactions on Visualization and Computer Graphics}, 2016.
\newblock \url{http://vgl.serc.iisc.ernet.in/felix/index.html}.

\bibitem{sousbie11}
T.~Sousbie.
\newblock The persistent cosmic web and its filamentary structure: Theory and
  implementations.
\newblock {\em Royal Astronomical Society}, 2011.
\newblock \url{http://www2.iap.fr/users/sousbie/web/html/indexd41d.html}.

\bibitem{szymczak13}
A.~Szymczak.
\newblock Hierarchy of stable morse decompositions.
\newblock {\em IEEE Transactions on Visualization and Computer Graphics},
  19(5):799--810, 2013.

\bibitem{tenenbaum00}
J.~B. Tenenbaum, V.~de~Silva, and J.~Langford.
\newblock A global geometric framework for nonlinear dimensionality reduction.
\newblock {\em Science}, 2000.

\bibitem{thomas14}
D.~M. Thomas and V.~Natarajan.
\newblock Multiscale symmetry detection in scalar fields by clustering
  contours.
\newblock {\em IEEE Transactions on Visualization and Computer Graphics (Proc.
  of IEEE VIS)}, 2014.

\bibitem{ttk17}
J.~Tierny, G.~Favelier, J.~A. Levine, C.~Gueunet, and M.~Michaux.
\newblock The {T}opology {T}ool{K}it.
\newblock {\em IEEE Transactions on Visualization and Computer Graphics (Proc.
  of IEEE VIS)}, 2017.
\newblock \url{https://topology-tool-kit.github.io/}.

\bibitem{TurnerMMH14}
K.~Turner, Y.~Mileyko, S.~Mukherjee, and J.~Harer.
\newblock Fr{\'{e}}chet means for distributions of persistence diagrams.
\newblock {\em Discrete {\&} Computational Geometry}, 2014.

\bibitem{vonLuxburg2007}
U.~von Luxburg.
\newblock A tutorial on spectral clustering.
\newblock In {\em Statistics and Computing}, 2007.

\bibitem{WhitakerMK13}
R.~T. Whitaker, M.~Mirzargar, and R.~M. Kirby.
\newblock Contour boxplots: {A} method for characterizing uncertainty in
  feature sets from simulation ensembles.
\newblock {\em IEEE Transactions on Visualization and Computer Graphics (Proc.
  of IEEE VIS)}, 2013.

\bibitem{wickelmaier2003introduction}
F.~Wickelmaier.
\newblock An introduction to mds.
\newblock Technical report, Aalborg University, 2003.

\end{thebibliography}
\end{document}